\documentclass[twocolumn]{aastex631}

\usepackage{pbox, cellspace}
\usepackage{hyperref}
\renewcommand{\arraystretch}{2}

\usepackage[T1]{fontenc}
\usepackage{graphicx}	
\usepackage{amsmath}	
\usepackage{amssymb}	
\usepackage{wrapfig}
\usepackage{natbib}
\usepackage{xcolor}
\usepackage{ulem}
\usepackage{soul}
\definecolor{blazeorange}{rgb}{1.0, 0.4, 0.0}
\definecolor{seagreen}{rgb}{0.18, 0.55, 0.34}
\definecolor{rufous}{rgb}{0.66, 0.11, 0.03}
\definecolor{royalfuchsia}{rgb}{0.79, 0.17, 0.57}
\definecolor{scarlet}{rgb}{1.0, 0.13, 0.0}
\definecolor{royalpurple}{rgb}{0.47, 0.32, 0.66}
\definecolor{darkblue}{rgb}{0, 0, 0.66}

\DeclareRobustCommand{\VAN}[3]{#2}
\let\VANthebibliography\thebibliography
\def\thebibliography{\DeclareRobustCommand{\VAN}[3]{##3}\VANthebibliography}

\hypersetup{linkcolor=blue,citecolor=blue,filecolor=cyan,urlcolor=magenta}
\usepackage{newtxtext,newtxmath}
\usepackage{threeparttable}
\usepackage{mathtools}

\def\d{\mathrm{d}}

\newcommand{\lta}{\lower 2pt \hbox{$\, \buildrel {\scriptstyle <}\over {\scriptstyle \sim}\,$}}
\newcommand{\gta}{\lower 2pt \hbox{$\, \buildrel {\scriptstyle >}\over {\scriptstyle \sim}\,$}}
\def\approxprop{%
  \def\p{%
    \setbox0=\vbox{\hbox{$\propto$}}%
    \ht0=0.6ex \box0 }%
  \def\s{%
    \vbox{\hbox{$\sim$}}%
  }%
  \mathrel{\raisebox{0.7ex}{%
      \mbox{$\underset{\s}{\p}$}%
    }}%
}

\begin{document}

\title{Can repeating and non-repeating FRBs be drawn from the same population?}

\correspondingauthor{Paz Beniamini}
\email{pazb@openu.ac.il}

\author[0000-0001-7833-1043]{Paz Beniamini}
\affiliation{Department of Natural Sciences, The Open University of Israel, P.O Box 808, Ra'anana 4353701, Israel}
\affiliation{Astrophysics Research Center of the Open university (ARCO), The Open University of Israel, P.O Box 808, Ra'anana 4353701, Israel}
\affiliation{Department of Physics, The George Washington University, 725 21st Street NW, Washington, DC 20052, USA}

\author{Pawan Kumar}
\affiliation{Department of Astronomy, University of Texas at Austin, Austin, TX 78712, USA}

\begin{abstract}
Do all Fast Radio Burst (FRB) sources repeat? We present evidence that FRB sources follow a Zipf-like distribution, in which the number density of sources is approximately inversely proportional to their burst rate above a fixed energy threshold-even though both the burst rate and number density span many orders of magnitude individually. We introduce a model-independent framework that predicts the distribution of observed fluences and distances, and repetition rates of an FRB population based on an assumed burst rate distribution per source. Using parameters derived directly from observations, this framework simultaneously explains several key features of the FRB population: (i) The observed ratio of repeaters to apparent non-repeaters; (ii) The much lower ratio of apparent non-repeaters to the total number of Soft Gamma Repeater (SGR) sources within the observable Universe; And (iii) the slightly smaller average distances of known repeaters compared to non-repeaters. We further explore how survey parameters, such as radio sensitivity and observation time, influence these statistics. Notably, we find that the fraction of repeaters rises only mildly with improved sensitivity or longer exposure. This weak dependence could be misinterpreted as evidence that not all FRBs repeat. Overall, our results support the idea that a single population-likely magnetars-can account for the full observed diversity of FRB activity, from very inactive FRB sources like SGR 1935+2154 to the most active repeaters.
\end{abstract}


\section{Introduction} 
\label{sec:intro}
Fast radio bursts (FRBs) exhibit a striking diversity in their observable properties. The isotropic-equivalent energies of extragalactic FRBs with known redshifts span over nine orders of magnitude - from as low as $\sim 6\times 10^{23}\mbox{erg Hz}^{-1}$ in FRB 20200120E \citep{Nimmo2023} to as high as $\gtrsim 6\times 10^{32}\mbox{erg Hz}^{-1}$ in FRB 20220610A \citep{Ryder2023}.
Even individual repeaters can show large variability: for example, bursts from FRB 20201124A cover about five orders of magnitude in energy \citep{Zhang20201124A,2022MNRAS.512.3400K}. 
FRB sources also differ greatly in their apparent activity rate \citep{MBSM2020}. Some, like FRBs 20121102A (aka R1) and 20201124A, have been observed to emit dozens of bursts in a single hour \citep{2021Natur.598..267L,Zhang20201124A} while others - such as the Galactic magnetar SGR 1935+2154 - have only produced a single energetic burst with in an observed period of over several years. As discussed in \S \ref{sec:varyrate}, this contrast becomes even more pronounced when comparing sources at similar energies, due to the steep energy dependence of burst rates.
A third axis of variation is the inferred number density of sources. For instance, \cite{LBK2022} showed that the volumetric density of highly active extragalactic FRBs like 20180916B and 20121102A is orders of magnitude lower than that of low-activity sources such as SGR 1935+2154 or FRB 20200120E in the nearby M81 galaxy. Given these dramatic differences - in energy output, repetition rate, and source density - it remains an open question whether all FRBs can be unified under a single physical framework, and if so, what underlying parameters account for their apparent diversity.

Perhaps the most striking of the apparent dichotomies, is the observational separation between sources seen only once thus far (``non-repeaters") and those seen multiple times (``repeaters"). The most reliable way to estimate the intrinsic fraction of repeating sources is to consider a blind survey mission such as CHIME. In the first 20 months of its operation, CHIME detected 46 repeating sources (and 14 more candidates), consisting a fraction of $2.6_{-2.59}^{+2.9}\%$ of sources in their sample \citep{CHIME2023repeaters}. The lack of bimodality in observed burst rates per source between the population of repeaters and (as of yet) non-repeaters detected by CHIME (with observed rates varying continuously from a few per hour to $\lesssim 10^{-3}\mbox{ hr}^{-1}$, \citealt{CHIME2023repeaters}), suggests all FRB sources might repeat, and that the low observed fraction is just a reflection of the current sensitivity. 
The finding that repeaters in this sample have, on average, slightly lower extragalactic DMs than non-repeaters ($\sim430\mbox{ pc cm}^{-3}$ as compared with $\sim 600\mbox{ pc cm}^{-3}$), could simply be a selection effect - repeaters are by the nature of their identified repetition, likely to be slightly closer than non-repeaters. This does not require intrinsically distinct populations. Similarly, while there are some small differences between repeater and non-repeater host galaxies (the former being slightly less massive, and having lower metallicity, \citealt{Sharma2024}), these differences are not statistically meaningful and might in addition be subject to further selection biases. 

A less trivial difference between repeaters and non-repeaters is with regards to their spectro-temporal properties. \cite{Scholz2017,Fonseca2020,Pleunis2021B} have found that FRBs from repeaters are longer in duration and spectrally narrower than non-repeaters. As shown by \cite{Metzger2022}, such an apparent bi-modality can be a natural consequence of variation in a single intrinsic parameter, controlling the rate of central frequency drift. When the drift is slow, bursts maintain their intrinsically narrow (time-resolved) spectrum and remain within the detector band for longer. Alternatively, when the drift is fast, the peak frequency sweeps so quickly, that the bursts are temporally unresolved, and their spectra integrated over one temporal bin of the detector appear as wide power laws (PLs). The origin of the frequency drift rate, might have a geometrical explanation, whereby active repeaters are magnetars with the spin, magnetic and line of sight directions all roughly aligned and non-repeaters are preferentially misaligned sources \citep{BK2025}. This interpretation can also help explain the discrepancies mentioned above regarding apparent activity rates and source densities.

A different way to assess the possibility of a shared origin of repeaters and non-repeaters, is to monitor over a long time known, highly active, repeater sources. The energy distribution per source is found (at least by some studies) to be consistent with the energy distribution of non-repeaters \citep{Wang2017,EnergyaandWaiting121102}. \cite{Kirsten2024} have observed FRB 20201124A for over 2000 hours, using 25-32 m class telescopes. They have found that the slope of the burst energy distribution varies both as a function of time (sporadically) and energy (becoming shallower at higher energies). The higher energy slope is consistent with that of non-repeating sources and the time variation helps explain the apparent large range of repeating rates between sources. Similar results (utilizing 1500 hours of observations) were found for another repeater, FRB 20220912A \citep{Ould-Boukattine2024}.

In this paper, we set out to explore whether the activity rates, source densities, and energies of FRB repeaters / non-repeaters can originate from the same underlying population. Our approach is primarily analytic (backed by numerical calculations) and is independent of the nature of FRB sources.
We begin, in \S \ref{sec:Singlesource} by exploring the simplest hypothesis supporting a unified FRB origin: that there is a single source type powering all observed FRBs. Then, in \S \ref{sec:varyrate} (and \S \ref{sec:correlationb}) we expand the model to allow for variable intrinsic rates and typical burst energies. The results are then compared with observations in \S \ref{sec:Infer} and implications for magnetar sources are discussed in \S \ref{sec:magnetar}. We conclude in \S \ref{Sec:conclude}.

\section{Single source rate}
\label{sec:Singlesource}
Consider a single population of FRB progenitors, with a volumetric source density $n_{\rm src}$ and an energy dependent\footnote{If the bursts are beamed, the quoted energies should be understood as isotropic-equivalent values. When the emission is isotropically distributed over time and the beaming fraction is constant and energy-independent, the observed burst rate per source is simply reduced by a constant factor $f_b<1$ relative to the intrinsic rate (see, e.g., \citealt{Beniamini2025}). However, if the effective beaming varies between sources - for example, due to rotation of a localized emission region (e.g., a polar cap) that only intersects the observer's line of sight for part of the spin period - then the observed rates can vary significantly depending on the alignment of the spin axis and line of sight (see Appendix A.2 in \citealt{BK2025}). This naturally leads to a spread in inferred burst rates and corresponding source number densities, as discussed in \S \ref{sec:varyrate}.} rate of bursts per source \footnote{While some studies in the literature use the same notation as in Eq. \ref{FRB-rate}, others define instead $\mathcal{R}(>E_{\nu})\propto E_{\nu}^{\gamma}$. One should use caution when comparing the values of $\gamma$ from different works.}, 
\begin{equation}
\mathcal{R}(>E_{\nu})=\mathcal{R}_{\rm 0,*}(E_{\nu}/E_{\rm *})^{1-\gamma}
\label{FRB-rate}
\end{equation}
for $E_{\rm min}<E_{\nu}<E_{\rm c}$ and $\gamma>1$, where $E_{\nu}$ is the energy per unit frequency, $E_{\rm c}$ is the maximum burst energy per source\footnote{Some authors have fit the burst distribution with a functional form of a Schechter function: $d\mathcal{R}/dE_{\nu}\propto (E_{\nu}/E_{\rm c})^{-\gamma}e^{-E_{\nu}/E_{\rm c}}$. In that case $E_{\rm c}$ is the energy marking the transition from the PL distribution to the exponential cut-off.} and, in anticipation of \S \ref{sec:varyrate}, we use two subscripts for the normalization rate, $\mathcal{R}_{\rm 0,*}$, the first referring to the activity level of the source (0 denoting the most common and inactive sources or the only sources in the case of a single source activity level explored in this section), and the second to the burst energy at which the normalization is taken. $E_*$ is an energy used for the normalization of the rate per source. Throughout the majority of this work we choose $E_{\rm *}=E_{\rm c}$ (an exception to this is discussed in \S  \ref{sec:correlationb}).

At the simplest level, a large sky FRB survey can be described by its limiting (spectral) fluence $\Phi_{\rm lim}$ and by the time, $T$, it spends observing each point on the sky \footnote{We first consider a uniform $T$ value across the sky; we provide a generalized treatment in \S \ref{sec:MonteCarlo}. During the data collection period for the 1st CHIME catalog, the median exposure time was $\sim \! 20$\,hrs \citep{CHIME_1st_cat}, which we adopt as a reference value when using a fixed $T$.}. While low energy bursts are more common per source (Eq. \ref{FRB-rate}), such bursts cannot be detected to large distances by a fluence-limited survey. At the same time the number of sources increases rapidly with distance. The result is a competition between the detectability of low energy and close-by sources and high energy far-away sources. For $\gamma<2.5$, Euclidean geometry and a uniform volumetric density of sources \footnote{A verification of our results using all relevant cosmological corrections and not assuming a uniform volumetric source density is presented in \S \ref{sec:MonteCarlo}.}, the competition is won by the far-away sources (see, e.g. \citealt{LuPiro19}). That is, most detected bursts are from large cosmological distances. This holds up to a distance $r_z$ (roughly corresponding to the proper distance up to $z\approx 2$) beyond which the number of sources increases much more slowly, due to a significant departure from Euclidean geometry (as well as a likely intrinsic reduction in the source density, at eras before the peak of star formation history). The combination of $r_z$ and $\Phi_{\rm lim}$ yields the typical energy of detectable non-repeater bursts $E_{\rm lim}\equiv 4\pi r_z^2\Phi_{\rm lim}$ (those that are close to the threshold fluence and arise from distant sources).
However, the conclusion that most detected bursts have an energy of order $E_{\rm lim}$ does not necessarily hold for active repeaters, observed to repeat $k\gg 1$ times during the survey's operation. This is because, if most bursts from sources seen $k$ times have such an energy, then this corresponds to a single mean number of repetitions per source during the observed period, $\lambda_{{\rm lim}}\equiv \lambda_{\rm E_{\rm lim}}=\mathcal{R}(>E_{\rm lim})T$. 
Assuming Poisson statistics, the probability that a source will repeat $k$ times during $T$ is $P(k)=\lambda^k e^{-\lambda}/k!$ \footnote{Several studies reported non-Poissonian burst arrival time statistics from prolific repeaters, with bursts tending to cluster in time \citep{Oppermann2018,Wang2018,Oostrum2020}. However, \cite{Cruces2021,Jahns2023} showed that once short separations (on the order of tens of ms, likely corresponding to non-independent sub-bursts) are excluded and periodically modulated activity is accounted for, the remaining distribution is consistent with a Poisson process. Since deviations from Poissonian behavior would require additional model parameters, we adopt a Poissonian model here to avoid overfitting. This assumption can be revisited if future observations provide strong evidence for non-Poissonian statistics.}. Therefore, $\lambda/2$ is approximately the ratio between repeaters (sources observed at least twice) and non-repeaters and calibration with the observed ratio leads to $\lambda_{\rm lim}\approx 0.05$.  $P(k)$ then becomes vanishingly small for large $k$. For example $P(k\geq 10)/P(k\geq 1)\approx \lambda^{10}/10!\sim 4\times 10^{-20}$. The fact that CHIME has detected multiple sources with more than 10 repetitions clearly means that such a picture is too simplistic and that the distance and / or intrinsic rate per source must be different for sources detected multiple times. The first point is addressed below and the second in \S \ref{sec:varyrate}.

Under the assumption of a single source type, the value of $\lambda_{\rm lim}\approx 0.05$ calibrated by the repeater / non-repeater ratio can be used to deduce the rate per source above $E_{\rm lim}$,
\begin{eqnarray}
    &\mathcal{R}(>E_{\rm lim})\!\approx \!2.5\times10^{-3} \left(\frac{20\mbox{ hr}}{T}\right) \left(\frac{\lambda_{\rm lim}}{0.05}\right)\mbox{ hr}^{-1}\nonumber\\
    &E_{\rm lim}\approx 1.6 \times 10^{32}\left(\frac{\Phi_{\rm lim}}{5\mbox{Jy ms}}\right)\mbox{ erg Hz}^{-1} .
\end{eqnarray}
For numerical estimates, we adopt $\Phi_{\rm lim}=5$Jy ms, which is the 95$\%$ completeness threshold applicable to the 1st CHIME catalog.
As $\lambda_{E_{\nu}}$ increases with decreasing $E_{\nu}$, one can define a critical energy, $E_{\rm all,k}$ such that $P(>E_{\rm all,k})\approx 1$ (this energy is approximately where the Poisson probability distribution function peaks, or $\lambda(>E_{\rm all,k})=k$). Plugging $E_{\rm all,k}$ into Eq.\ref{FRB-rate}, multiplying by $T$ and rearranging, we get
\begin{equation}
    E_{\rm all,k}\!=\!E_{\rm c} \left(\frac{\lambda_{\rm c}}{k}\right)^{1\over \gamma-1}\!=\!E_{\rm lim} \left(\frac{\lambda_{\rm lim}}{k}\right)^{1\over \gamma-1}\!<\!E_{\rm lim}
\end{equation}
where $\lambda_{\rm c}\!\equiv \!\lambda_{E_{\rm c}}$ and (since by definition $k\!\geq\! 1$) the last inequality necessarily holds for $\lambda_{\rm lim}\!<\!1$. 

We define the fluences $\Phi_{\rm c}\!=
\!E_{\rm c}/(4\pi r_z^2)$ and
$\Phi_{\rm all,k}\!=\!E_{\rm all,k}/(4\pi r_z^2)\!\approx \!(\lambda_{\rm c}/k)^{1\over \gamma-1} \Phi_{\rm c}$ which are respectively the fluences corresponding to bursts with energy $E_{\rm c}$ and $E_{\rm all,k}$ as viewed from a distance $r_z$.
Below $\Phi_{\rm all,k}$, roughly all sources in the Universe are observed by the survey $\geq k$ times. Taking $E_{\rm c}\!\approx\! 3\!\times\! 10^{33}\mbox{erg Hz}^{-1}, \gamma\!\approx\! 1.7$ as inferred from FRB energy distribution studies calibrated by observations \citep{LuPiro19,James2022,Hashimoto2022,Shin2023,Lin2024,Gupta2025}
and recalling that $\lambda_{\rm c}=\lambda_{\rm lim} (E_{\rm c}/E_{\rm lim})^{1-\gamma}$ (Plugging $E_{\nu}=E_{\rm c}$ in Eq. \ref{FRB-rate} and multiplying by $T$),
we find $\lambda_{\rm c}\!\approx\! 7\!\times\! 10^{-3}, \Phi_{\rm c}\!\approx \!100\mbox{Jy ms}, \Phi_{\rm all,1}\!\approx \!0.07\mbox{Jy ms}$, so we expect the ordering $\Phi_{\rm all,1}\!\ll\!\Phi_{\rm lim}\!\ll\!\Phi_{\rm c}$ to hold for CHIME.

Denote by $N_k$ the number of sources seen to repeat $k$ times above a fluence $\Phi_{\rm lim}$.
In the energy limited regime ($\Phi_{\rm lim}\!>\!\Phi_{\rm c}$), $N_k\! \propto \lambda_{\rm lim}^k r^3\!\propto (\Phi_{\rm lim} r^2)^{k(1-\gamma)}r^3\! \propto r^{2k(1-\gamma)+3}\propto E^{k(1-\gamma)+3/2}$ (where in the last transition we have used $E\propto \Phi_{\rm lim} r^2$). This shows there is a critical $k_{\gamma}=3/(2(\gamma-1))$ and that the observed distribution of distances / energies will depend on $k$ relative to $k_{\gamma}$.
Consider first $k\!>\!k_{\gamma}$. 
In this range, $N_k$ decreases with $E$, and bursts with lower energy are detected more commonly than those with larger energies. This holds down to the energy $E_{\rm all,k}$, below which all sources are detected at least $k$ times. Therefore at $\Phi<\Phi_{\rm all,k}$, all sources are observable. For a uniform density of sources, and Euclidean geometry, their cumulative distance distribution is $N_k(r)=(r/r_z)^3$ and the median distance is $r=0.5^{1/3} r_z\approx 0.8r_z$. At larger fluences, the most distant sources are missed. Fixing the energy of the bursts at $E_{\rm all,k}$ as explained above, we see that the maximum detectable distance becomes $r(\Phi)=(E_{\rm all,k}/(4\pi \Phi))^{1/2}=r_z(\Phi_{\rm all,k}/\Phi)^{1/2}$. Overall, 
\begin{eqnarray}
\label{eq:rtypkgtr}
& r_{\rm typ}(>\!\Phi;k\!>\!k_{\gamma})\!\approx\!0.8 r_z  \left\{ \begin{array}{ll}\!1 & ; \Phi\!<\!\Phi_{\rm all,k} \\
\!\sqrt{\frac{\Phi_{\rm all,k}}{\Phi}} & ; \Phi>\Phi_{\rm all,k}.
\end{array}\right. 
\end{eqnarray}
Instead, for $k<k_{\gamma}$, the most energetic bursts, with $E_{\nu}\approx E_{\rm c}$ are the ones detected most often.
If $\Phi<\Phi_{\rm c}$, then energetic bursts with $E_{\rm c}$ can be seen from any $r<r_z$, while for $\Phi>\Phi_{\rm c}$ the typical distances are limited by the maximum distance to which $E_{\rm c}$ can be detected
\begin{eqnarray}
\label{eq:rtypksml}
& r_{\rm typ}(>\!\Phi;k\!<\!k_{\gamma})\!\approx\!0.8 r_z  \left\{ \begin{array}{ll}\!1 & ; \Phi\!<\!\Phi_{\rm c} \\
\!\sqrt{\frac{\Phi_{\rm c}}{\Phi}} & ; \Phi>\Phi_{\rm c} .
\end{array}\right. 
\end{eqnarray}
Put simply, assuming a single source rate and for $\Phi_{\rm all,k}<\Phi<\Phi_{\rm c}$, very active repeaters will be dominated by nearby sources (bottom part of Eq. \ref{eq:rtypkgtr})\footnote{In practice, the bottom lines of Eqns. \ref{eq:rtypkgtr}, \ref{eq:rtypksml} hold only as long as there is at least one source at this distance. We define the typical distance within which there is one source as $n_{\rm src}(4\pi/3)r_1^3=1\to r_1=(3/(4\pi n_{\rm src}))^{1/3}$ (where $n_{\rm src}$ is the source density). In general, sources will reside at $r\!\approx \!\max(r_{\rm typ},r_1)$.}, while non-repeaters by distant ones (top part of Eq. \ref{eq:rtypksml}).
The fluence distribution of sources repeating $k$ times is $N_{\rm k}(>\Phi)\propto \Phi^{k(1-\gamma)}$. For $k>k_{\gamma}$ the latter is steeper than the result for a constant energy of bursts in Euclidean geometry, $N_{\rm k}(>\Phi)\propto \Phi^{-3/2}$, and therefore it is sub-dominant (the contribution of less energetic $E_{\nu}\approx E_{\rm all,k}$, but closer, sources outpaces that of rarer but more energetic events that are visible to larger distances).

To summarize, we have\footnote{The simplified broken PL approximation holds well for large $k$. For lower values of $k$ there is significant curvature in the shape of $N_{\geq k}(>\!\Phi)$ around $\Phi_{\rm all,k}$ and as a result the asymptotic normalization is larger than given by Eq.\ref{eq:NkPhi}.}
\begin{eqnarray}
\label{eq:NkPhi}
& N_{\geq k}(>\!\Phi)\!=\! \sum_k^\infty\int_0^{\min[r_z,\sqrt{\frac{E_{\rm c}}{4\pi \Phi}}]}dr \frac{e^{-\lambda_E}\lambda_E^k}{k!}n_{\rm src} 4\pi r^2 \nonumber \\
    &\!\approx\!N_z \! \left\{ \begin{array}{ll}\!1 \!&\! \Phi\!<\!\Phi_{\rm all,k}\!<\!\Phi_{\rm c} \\
\!\left(\frac{\Phi_{\rm all,k}}{\Phi}\right)^{\min[k(\gamma\!-\!1),\frac{3}{2}]} \!& \! \Phi_{\rm all,k}\!<\!\Phi\!<\!\Phi_{\rm c} \\ \!\left(\frac{\Phi_{\rm all,k}}{\Phi_{\rm c}}\right)^{\min[k(\gamma\!-\!1),\frac{3}{2}]} \!\left(\frac{\Phi_{\rm c}}{\Phi}\right)^{\frac{3}{2}} \!&\! \Phi_{\rm all,k}\!<\!\Phi_{\rm c}\!<\!\Phi\ .
\end{array}\right.   
\end{eqnarray}
where $N_z=n_{\rm src}\frac{4\pi}{3}r_z^3$ is the number of sources up to $r_z$.
Similarly, the ratio between repeaters and non-repeaters is given by
\begin{eqnarray}
\label{eq:repratio}
& \frac{N_{\geq k}(>\Phi)}{N_{\geq 1}(>\Phi)} \approx  \\ & \scalebox{0.75}{$ \left\{ \begin{array}{ll}\!1 \!&\! \Phi\!<\!\Phi_{\rm all,k} \\
\! \max\left[\left(\frac{\Phi_{\rm all,k}}{\Phi}\right)^{k(\gamma\!-\!1)},\left(\frac{\Phi_{\rm all,k}}{\Phi}\right)^{\frac{3}{2}}\right] \!&\! \Phi_{\rm all,k}\!<\!\Phi\!<\!\Phi_{\rm all,1} \\
\!\max\left[\left(\frac{\Phi_{\rm all,k}}{\Phi}\right)^{(k\!-\!1)(\gamma\!-\!1)},\left(\frac{\Phi_{\rm all,k}}{\Phi}\right)^{\frac{5}{2}\!-\!\gamma}\right] \!& \! \Phi_{\rm all,1}\!<\!\Phi\!<\!\Phi_{\rm c} \\ \!\max\left[\left(\frac{\Phi_{\rm all,k}}{\Phi_{\rm c}}\right)^{(k\!-\!1)(\gamma\!-\!1)},\left(\frac{\Phi_{\rm all,k}}{\Phi_{\rm c}}\right)^{\frac{5}{2}\!-\!\gamma}\right] \!&\! \Phi_{\rm c}\!<\!\Phi\ .
\end{array}\right. $} \nonumber 
\end{eqnarray}

\section{Varying activity rates}
\label{sec:varyrate}
We relax the assumption made in \S \ref{sec:Singlesource}, that all sources are identical, and assume, instead, that sources vary not only in their distances from us, but also in their activity rate. This is in addition to the energy dependence of burst rate per source, which is described by Eq. \ref{FRB-rate}. Specifically, at some fixed energy $E_*$, we assume that the rates of different sources follow a PL distribution of activity levels: 
 \begin{equation}
 \label{eq:a_def}
      P(\geq \mathcal{R}_{\rm *})=(\mathcal{R}_{\rm *}/\mathcal{R}_{\rm 0,*})^{-a}
 \end{equation}
where $\mathcal{R}_{\rm *}\equiv \mathcal{R}(E_{*})$ and as per \S \ref{sec:Singlesource}, without loss of generality, we consider below $E_*=E_{\rm c}$.

Crucially, the assumption of a PL rate distribution is directly supported by observations. \cite{LBK2022} estimated the number density of several empirically defined FRB sub-classes, based on the distance to their nearest known member and their burst rate at their typically observed energies. We extrapolate their rates using Eq. \ref{FRB-rate} (with $\gamma=1.7$ unless otherwise measured for the same source) to a common reference energy of $10^{28}\mbox{erg Hz}^{-1}$ from the closest available observed energy as well as to a central frequency of $\nu=600$MHz using $E_{\nu}\propto \nu^{\alpha}$ with $\alpha=-1.5$ \citep{Macquart2019,Shin2023}, unless otherwise measured from the same source. 
In addition, we include rate estimates for FRBs 20181030A, 20121102A, 20201124A, 20220912A, 20240114A, 20171020 and 20250316A - sources not covered in \citet{LBK2022} - based on the findings of \cite{Shannon2018,2021Natur.598..267L,Zhang20201124A,2023ApJ...955..142Z,CHIME2023repeaters,Kirsten2024,Ould-Boukattine2024,2024MNRAS.534.3331K,2025MNRAS.tmp..751T,2025arXiv250513297S,CHIME250316A,FAST240114A}. For SGR sources, we use updated rates from GReX observations \citep{Shila2025}. More details regarding the methodology of producing the figure are given in appendix \ref{sec:figure1_explained}.

\begin{figure*}
\centering
\vspace{-5mm}
\includegraphics[width=0.68\textwidth]{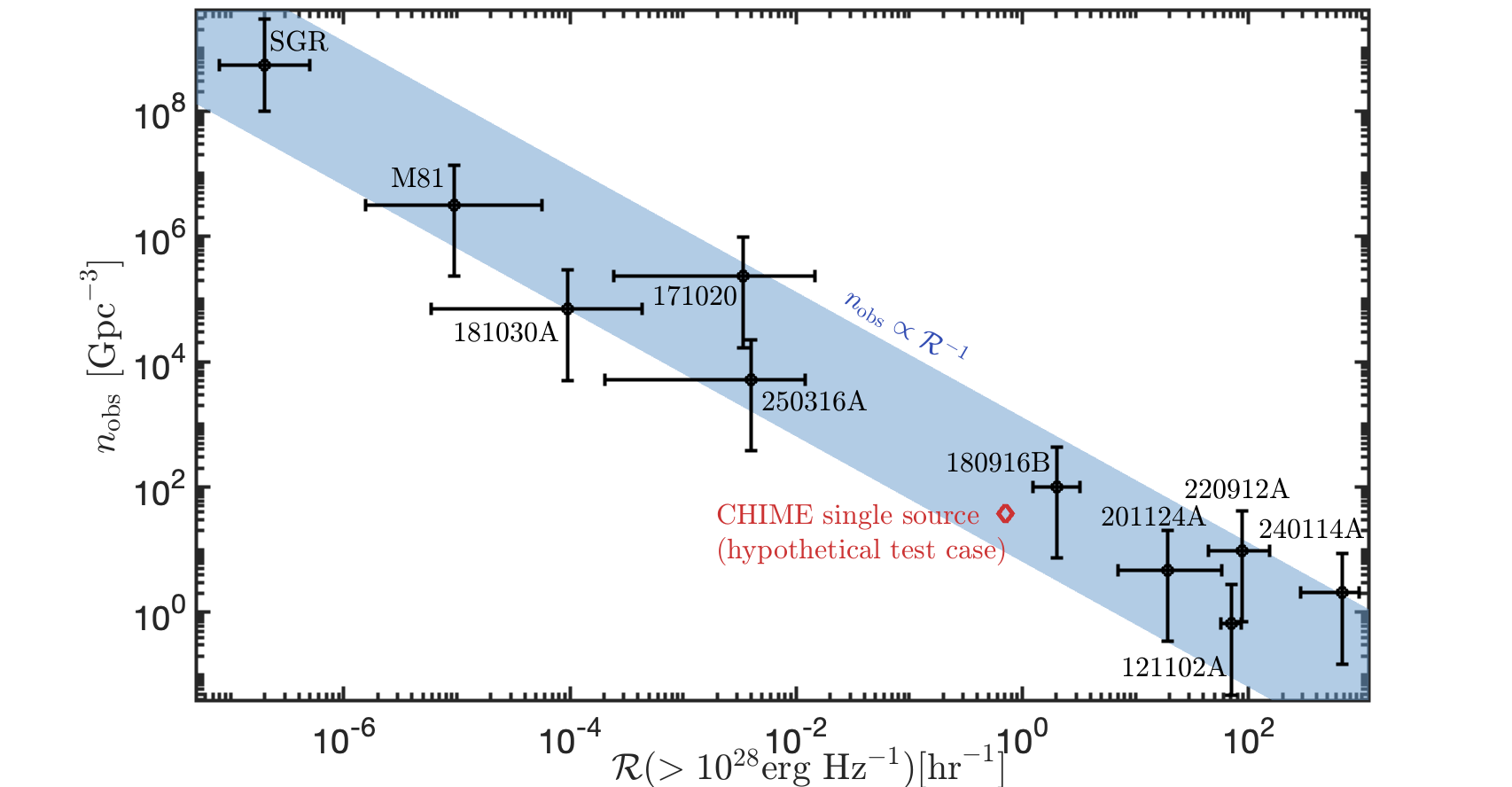}
\includegraphics[width=0.68\textwidth]{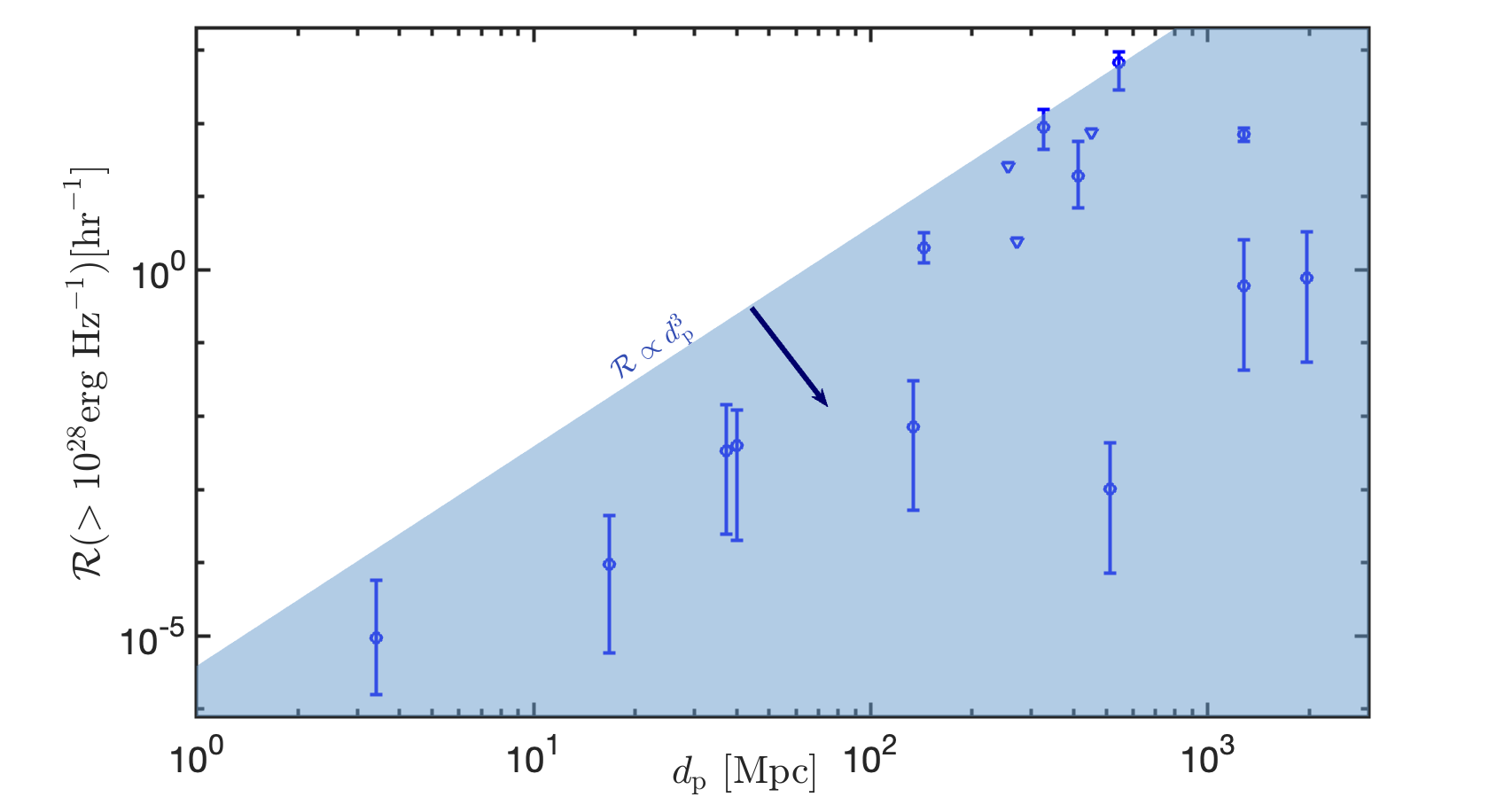}
\caption{Top: Estimated source number densities of different FRB sub-classes. These are plotted against the rate per source extrapolated to an intermediate common energy of $10^{28}\mbox{erg Hz}^{-1}$ using the measured $\gamma$ or $\gamma\!\approx \!1.7\!\pm\! 0.2$ when not available. Uncertainties mark the range corresponding to $90\%$ confidence limits. A diamond indicates the same properties for the hypothetical test case (disfavored by our analysis, see \S \ref{sec:figure1_explained} for details) that a single source type dominates CHIME data. Since the latter is dominated by non-repeaters, this also provides a rough estimate for the number density and rate of CHIME non-repeaters. We find that approximately $n_{\rm obs}\!\propto\! \mathcal{R}^{-1}$ over a factor of $\sim \! 10^9$ in density, suggesting that the large range between common/inactive and rare/active sources might be fit by a single PL distribution with roughly equal contributions at different logarithmic scales. Bottom: Rate per source as a function of (proper) distance. Results are shown for extragalactic FRB repeaters and non-repeaters with well constrained distances and rates. Downward facing triangles indicate $90\%$ upper limits. Consistency with the PL distribution of activity rates, $n_{\rm obs}\propto \mathcal{R}^{-a}$, requires that sources occupy the region below a boundary given by $\mathcal{R}\propto d_{\rm p}^{3/a}$ - i.e. within the blue shaded region.}
\label{fig:ratevsn}
\end{figure*}

The resulting distribution is shown in Fig. \ref{fig:ratevsn}.
While individual rate estimates are subject to uncertainties - and, as discussed in \S\ref{sec:intro}, some sources show temporal variability and deviate from a single PL energy distribution - the aim of Fig. \ref{fig:ratevsn} is to capture the overall population trends using a simple model that avoids over-fitting.
A striking feature of this plot is that the different FRB sub-classes span roughly nine orders of magnitude in number density and eight in burst rate (extrapolated to the fixed reference energy). Remarkably, the sources lie along a PL-like trend with $n_{\rm obs} \!\propto \!\mathcal{R}^{-1}$, in agreement with Eq. \ref{eq:a_def} and suggesting $a \approx 1$. As a comparison, we include in Fig.\ref{fig:ratevsn}, a point representing CHIME detected sources for the hypothetical test case (disfavored in this work) that there is a single activity rate per source underlying all sources (as per \S \ref{sec:Singlesource}). Details of this calculation are given in appendix \ref{sec:figure1_explained}.

The trend $n_{\rm obs}\propto \mathcal{R}^{-a}$ implies $\mathcal{R}\propto d_{\rm p}^{3/a}$ (where $d_{\rm p}$ is the proper distance). When extending the analysis to a larger group of sources, including those that are not the nearest known members of their population sub-class, one therefore expects known sources to span the `triangular' range in the distance - activity rate ($d_{\rm p}-\mathcal{R}$) plane up to the $\mathcal{R}\propto d_{\rm p}^{3/a}$ trend line. Indeed, this too is reproduced, as shown in the bottom panel of Fig. \ref{fig:ratevsn}. This supports the hypothesis of a single source population with a PL distribution of activity levels. That being said, we stress that the possibility of distinct multiple source populations cannot be ruled out - they are simply not required by these observational constraints.

An alternative, related possibility, is that sources that are more active at a particular energy, may also be able to produce more energetic bursts, i.e. that $\mathcal{R}_*$ correlates with $E_{\rm c}$. Such a possibility, and physical motivation for it, is described in appendix \S \ref{sec:correlationb}. As it unavoidably introduces additional apriori free parameters, we apply the principle of Occam's razor and focus primarily on the case with no such correlation. Future data will determine whether this complication is essential.

 \begin{figure*}
\centering
\vspace{-5mm}
\includegraphics[width=0.49\textwidth]{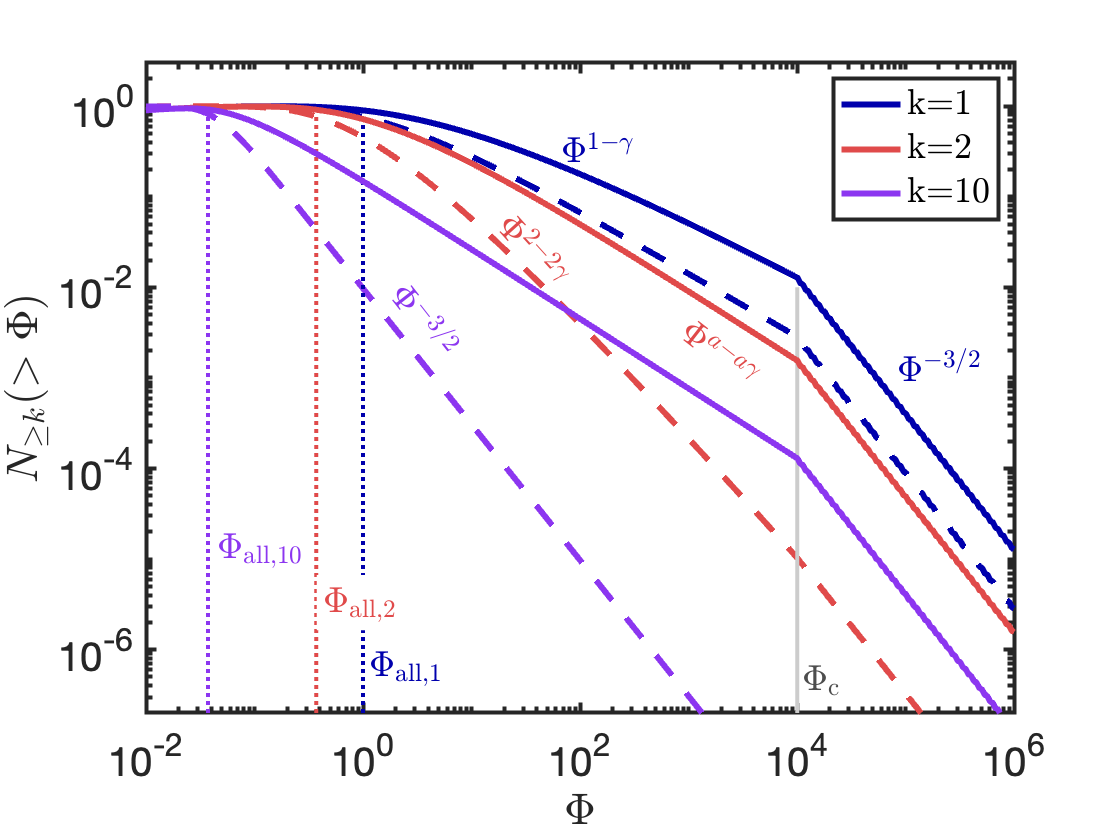}
\includegraphics[width=0.49\textwidth]{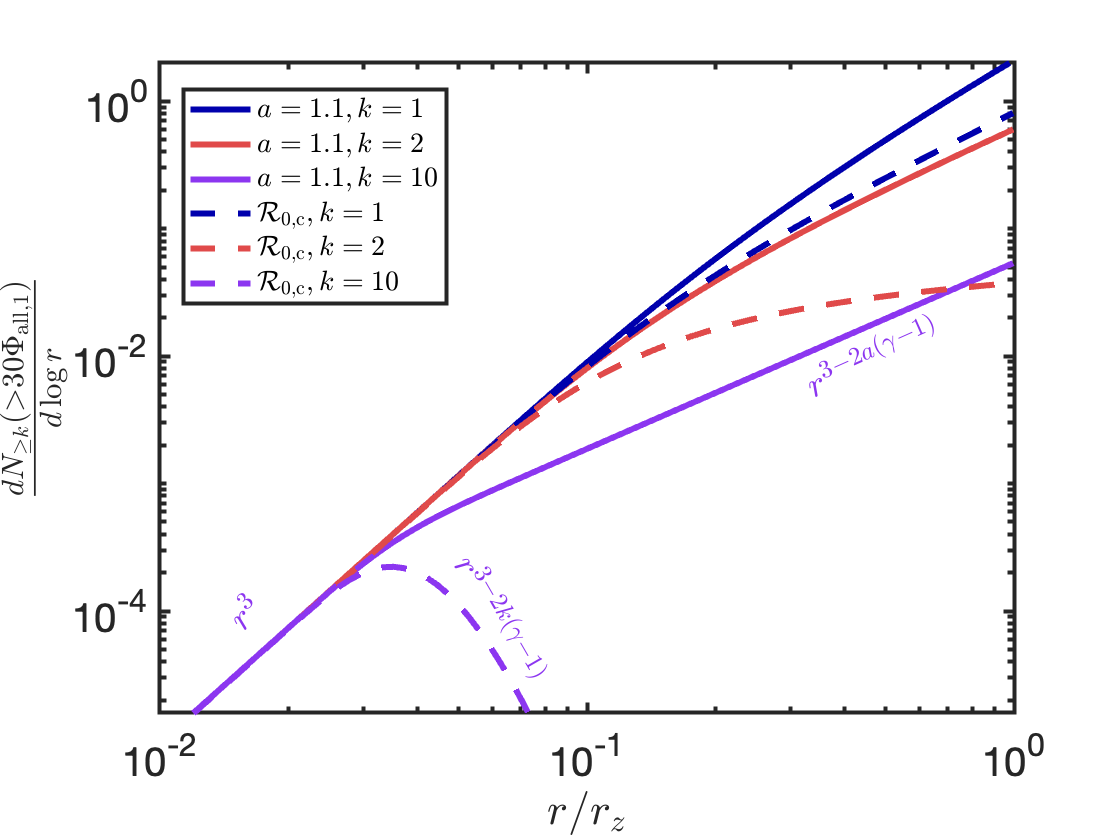}
\includegraphics[width=0.49\textwidth]{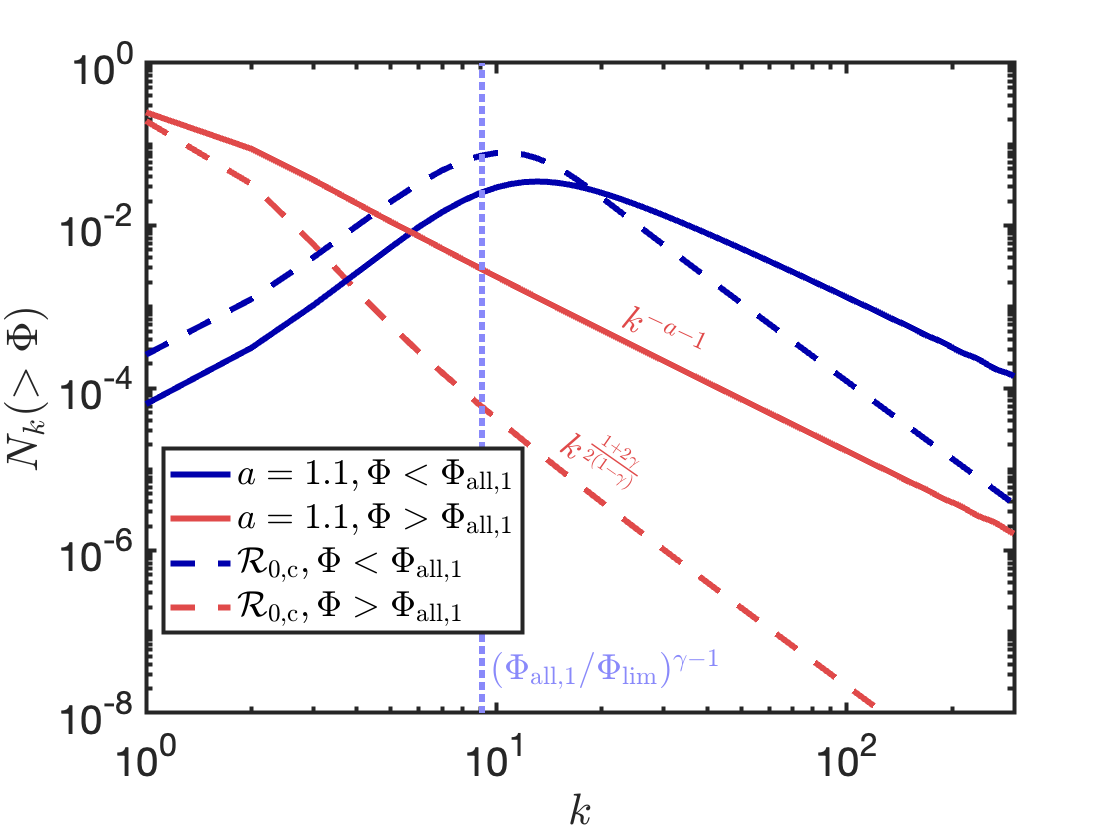}
\includegraphics[width=0.49\textwidth]{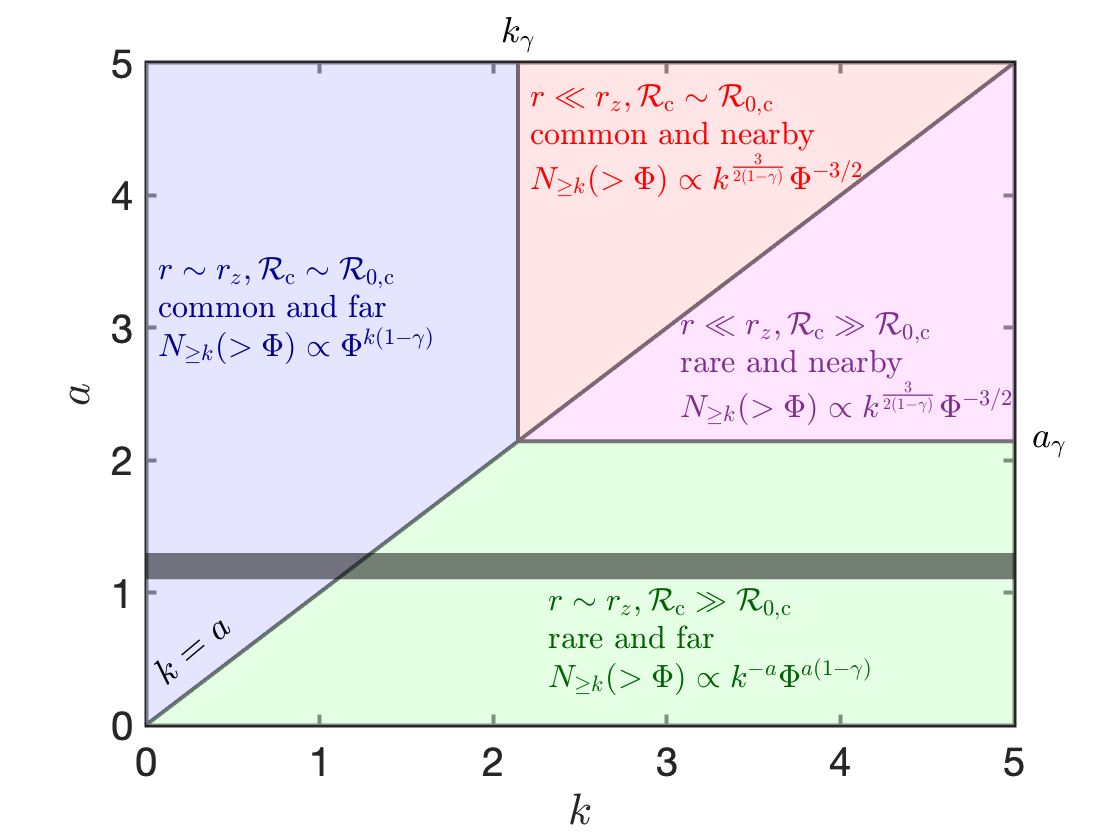}
\caption{Key results for the distributions of detectable sources. In all panels solid lines depict PL-distributed variable rates, $P(\geq \mathcal{R}_{\rm c})=(\mathcal{R}_{\rm *}/\mathcal{R}_{\rm 0,*})^{-a}$, and dashed lines a constant source rate. Top left: Number of sources above a given fluence with $\geq k$ repeats. For $k>k_{\gamma}$ and $a(\gamma-1)<3/2$, the distribution above $\Phi_{\rm all,k}$ is dominated by gradually rarer and more active sources. Top right: distance distribution of sources observed $\geq k$ times above a fluence $\Phi$. Bottom left: Number of sources observed k times as a function of k. For $\Phi<\Phi_{\rm all,1}$, the distribution peaks at k such that $\Phi_{\rm all,k}=\Phi$ ($k\approx (\Phi_{\rm all,1}/\Phi)^{\gamma-1}$; dotted line), while at larger $k$ the number declines as described in the text. Bottom right: Distances and typical observed activity rate in the parameter space of $a$ (describing the activity rate distribution) and $k$ (number of observed repetitions per source). We also list the dependence of $N_{\geq k}(>\Phi)$ on $k$ and $\Phi$ at the limit of large $k$ and for $\Phi_{\rm all,1}(\mathcal{R}_{\rm 0,c}.)<\Phi<\Phi_{\rm c}$. A shaded region marks the range of $a$ favored by the analysis in \S \ref{sec:Infer}.}
\label{fig:Nphivarrate}
\end{figure*}

The number of sources observable at $\Phi_{\rm all,k}\!<\!\Phi\!<\!\Phi_{\rm c}$ is given by $\frac{dN_{\geq k}}{d\log \Phi}=\Phi\frac{dN_{\geq k}}{d\mathcal{R}_{\rm *}}\frac{d\mathcal{R}_{\rm *}}{d\Phi}\propto \Phi^{a(1-\gamma)}$ where we have used the fact that $\Phi_{\rm all,k}\propto \mathcal{R}_{\rm *}^{\frac{1}{\gamma-1}}$. This scaling holds as long as it is flatter than the slope given by Eq. \ref{eq:NkPhi} (i.e. $a<k$ and $a<k_{\gamma}$). In this regime, common sources dominate at low fluences, and the typical intrinsic rate of an observable source increases with $\Phi$, until, at $\Phi > \Phi_{\rm c}$, the distribution becomes dominated by rare sources  with a burst rate of
$\mathcal{R}_{\rm *,1}=\mathcal{R}_{\rm 0,*}\lambda_{\rm c}(\mathcal{R}_{\rm 0,*})^{-1}>\mathcal{R}_{\rm 0,*}$.
Here, $\mathcal{R}_{\rm *,1}$ is the rate at which bursts with $E_{\rm *} = E_{\rm c}$ repeat at least once during the time $T$, which is equivalent to requiring $\Phi_{\rm all,1}(\mathcal{R}_{\rm c,1}) = \Phi_{\rm c}$.
It is instructive to consider also the distance distribution of observed sources, for which we estimate
\begin{eqnarray}
\label{eq:doublepartial}
    & \frac{\partial^2N_{\geq k}(>\Phi)}{\partial \log r \partial \log \mathcal{R}_{\rm *}}\!=\!4\pi n_{\rm src} r^3\!\left(\frac{\mathcal{R}_{\rm *}}{\mathcal{R}_{\rm 0,*}}\right)^{-a}\! P(\geq k| \lambda(\mathcal{R}_{\rm *},r,\Phi)) \nonumber \\
    & \propto r^{3-2k(\gamma-1)}\mathcal{R}_{\rm *}^{k-a}
\end{eqnarray}
where the last transition applies when $\lambda\ll 1$. Consider a constant activity rate. We see that while $\lambda\!\gtrsim\! 1$, $P\!\sim\! 1$ and so $\frac{\partial N_{\geq k}(>\Phi)}{\partial \log r }\!\propto\! r^3$ (all sources within limiting volume are visible). This lasts until $r\!\approx \!r_z (\Phi_{\rm all,k}/\Phi)^{1/2}$, beyond which only bursts with $E_{\nu}\!>\!E_{\rm all,k}$ are visible and $\frac{\partial N_{\geq k}(>\Phi)}{\partial \log r }\propto r^{3-2k(\gamma-1)}$ (which can have a positive or negative slope depending on $k/k_{\gamma}$). In particular, for $k\gg k_{\gamma}$, one finds that $r_{\rm typ}\propto k^{1\over 2(1-\gamma)}$ (Eq. \ref{eq:rtypkgtr}) and $N_{\geq k}(>\Phi)\!\propto \!r_{\rm typ}^3\!\propto\! k^{3\over 2(1-\gamma)}$ (or equivalently $N_k(>\Phi)\!\propto \! k^{1+2\gamma \over 2(1-\gamma)}$).

For a variable rate, the results remain qualitatively similar so long as $k\!\lesssim \!a$, such that the most common sources, with $\mathcal{R}_{\rm *}\!\approx \!\mathcal{R}_{\rm 0,*}$ dominate. Since $E_{\rm all,k}$ increases with $\mathcal{R}_{\rm *}$, for $k\!>\!a$ and $r\!>\! r_z \sqrt{\Phi_{\rm all,k}(\mathcal{R}_{\rm 0,*})/\Phi}$, sources with a rate such that $E_{\nu}\!=\!E_{\rm all,k}(\mathcal{R}_{\rm *})$ dominate. This leads to $\mathcal{R}_{\rm *}\!\propto\! r^{2(\gamma-1)}$ and using Eq. \ref{eq:doublepartial} to $\frac{\partial N_{\geq k}(>\Phi)}{\partial \log r }\propto r^{3-2a(\gamma-1)}$. The implication is that there is a critical value of $a\!\equiv \!a_{\gamma}\!=\!k_{\gamma}$, such that for $a\!<\!a_{\gamma}$ and $a<k$, most observed sources are distant and rare, while for $k\!>\!a\!>\!a_{\gamma}$, sources seen $k$ times are preferentially nearby and are associated with intrinsically active objects.
The results of the fluence and distance distributions with variable source rates, as well as the qualitative behavior in the $a-k$ parameter space, are illustrated in Fig. \ref{fig:Nphivarrate}. They can be divided into three cases:
\begin{enumerate}
    \item {\bf $a\!<\!a_{\gamma}\!=\!k_{\gamma}$ - } The least active of the observed sources (with $k\!<\!a$), are common (i.e. typical activity rate) and far ($r\!\sim \!r_z$). For larger $k\!>\!a$, the typical observed sources are instead dominated by rare (intrinsically highly active) and far sources. In this case $N_{\geq k}(>\!\Phi)\!\propto \!(\Phi/\Phi_{\rm all,k})^{a(1-\gamma)}\!\propto\! k^{-a}$ (or equivalently $N_{k}(>\!\Phi)\!\propto \!k^{-a-1}$). The number of sources with a given repetition number falls off slower than for the constant rate case.
    \item {\bf $a\!>\!a_{\gamma}\!=\!k_{\gamma}$ - } Sources observed with $k\!<\!k_{\gamma}$ are common (i.e. typical activity rate) and far ($r\!\sim \!r_z$). Sources seen $k_{\gamma}\!<\!k\!<\!a$ are common and nearby ($r\!\ll \!r_z$). Finally, the sources seen to repeat most often, with $k\!>\!a\!>\!k_{\gamma}$, are both rare and nearby. 
    \item {\bf Single activity rate - } $a\!\to \!\infty$. It is equivalent to the previous case, but where only the first two $k$ regimes are applicable.
\end{enumerate}

\section{Inferences from observations}
\label{sec:Infer}
With approximately one Milky Way-like galaxy per $100\mbox{Mpc}^{3}$\citep{Leroy2019} and 30 known active magnetars in the Milky Way \citep{MagnetarReview2017}, we can estimate the number density of FRB-producing magnetars in the local Universe as $10^7-3\times 10^8\mbox{Gpc}^{-3}$ (the range corresponding to either 1 or all magnetars in the Galaxy being potential FRB sources; for quantitative estimates below, we take the logarithmic mean). We use this to calculate the total number of FRB producing magnetars in the Universe, $N_{\rm src}\!\sim \!7\times \!10^{11}$, assuming that the magnetar density approximately tracks the star formation rate. At the same time, the number of sources seen by CHIME once or more during the period of data collection for the first CHIME catalog is only, $N_{\geq 1,\rm CH}^{\rm obs}=492$ and the number of sources observed at least twice (repeaters), $N_{\geq 2,\rm CH}^{\rm obs}=16$ \citep{CHIME_1st_cat}. Correcting for the fraction of the sky viewable by CHIME, $f_{\rm srv}\approx 0.5$ (i.e. the all-sky equivalent $N_{\geq 1,\rm CH}^{\rm all-sky}$ is given from the observed number $N_{\geq 1,\rm CH}^{\rm obs}$ via $N_{\geq 1,\rm CH}^{\rm all-sky}$=$N_{\geq 1,\rm CH}^{\rm obs}/f_{\rm srv}$), we see that
\begin{equation}
\label{eq:obsratio}
    \frac{N_{\rm \geq 1,CH}^{\rm all-sky}}{N_{\rm \geq 0,CH}^{\rm all-sky}}\approx 1.3\times 10^{-9\pm 0.74} \quad ; \quad \frac{N_{\rm \geq 2,CH}^{\rm all-sky}}{N_{\rm \geq 1,CH}^{\rm all-sky}}\approx 0.03
\end{equation}
FRB 20200428 had an energy of $\sim 10^{26}\mbox{ erg Hz}^{-1}$ \citep{STARE2020}. The rate of bursts of comparable energy from a given active magnetar in the Galaxy are constrained by over two years of observations by STARE2 and by $\sim 1.5$ years of observations by its successor GReX \citep{GReX,Shila2025}. These observations reduce the rate estimate based on STARE2 (e.g. \citealt{MBSM2020,LBK2022}) by a factor of several, resulting in $\mathcal{R}_{0}(\!>10^{26}\mbox{erg Hz}^{-1})\!\approx \!5\!\times \!10^{-6}\mbox{ hr}^{-1}$. From this we see that $\lambda_{0,*}\!\approx\! 10^{-4} (E_*/10^{26}\mbox{erg Hz}^{-1})^{1-\gamma}$. Observationally, we only have a lower limit on the maximum FRB energy from SGRs $E_*\!>\!10^{26}\mbox{ erg Hz}^{-1}$. If $E_*\!\gtrsim \!E_{\rm lim}$ then, $\lambda_{\rm lim}\!=\!\lambda_{\rm 0,lim}\!=\!\lambda_{0,*} (E_{\rm lim}/E_*)^{1-\gamma}\!\approx \!4.5\!\times \!10^{-9}$, which is of the same order as $N_{\geq 1}/N_{\geq 0}$. Is it possible to consistently explain the inferred values of $N_{\geq 1}/N_{\geq 0}$, $N_{\geq 2}/N_{\geq 1}$ and $\lambda_{\rm 0,lim}$?

Consider first the case of a single source type. Since $N_{\geq 2}/N_{\geq 1}\!\ll\! 1$, then $\lambda_{\rm lim}\!\ll \!1$, which using Eq. \ref{eq:repratio} leads to
\begin{equation}
\label{Eq:NSrc_sing}
    \frac{N_{\geq 1}}{N_{\geq 0}^{\rm sng}}\approx \lambda_{\rm lim}\approx \frac{4 N_{\geq 2}}{N_{\geq 1}}\approx 0.05
\end{equation}
which is too large by at least seven orders of magnitude (see Eq. \ref{eq:obsratio}) \footnote{Accounting for the curvature of $N_{\geq k}(>\Phi)$ as well as for the burst rate as a function of frequency, cosmological corrections, a more realistic survey observation time distribution (\S \ref{sec:MonteCarlo}) and a non-uniform density of sources, decreases $N_{\geq 1}/N_{\geq 0}$ by a factor of $\sim 26$ for a given  $N_{\geq 2}/N_{\geq 1}$. This is still many orders of magnitude too large to accommodate SGR-type sources and typical CHIME sources as part of a single distribution.}.

The difficulty of explaining both the observed ratio of repeaters to non-repeaters and the wide range of activity levels using a single type of source was first highlighted by \cite{MBSM2020}. That study proposed a bimodal population, consisting of rare but highly active sources and common but largely inactive ones, to account for the data. However, as shown in Fig. \ref{fig:ratevsn}, this bimodal model struggles to explain intermediate cases, such as the FRB associated with the M81 galaxy. Instead, the data suggest a continuous PL distribution, spanning from the most common, least active sources to the rarest, most active ones. It is this continuous framework that we explore below.

\begin{figure*}
\centering
\vspace{-5mm}
\includegraphics[width=0.49\textwidth]{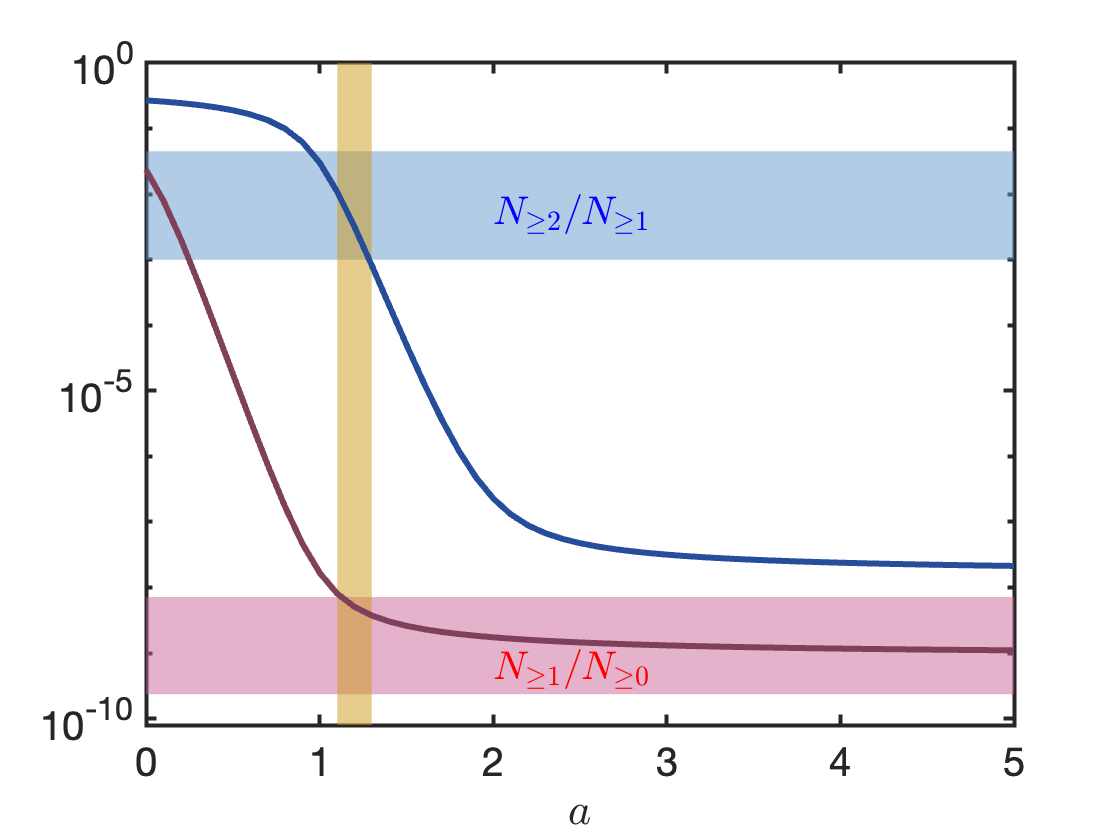}
\caption{Ratio of repeaters to non-repeaters (blue) and of non-repeaters to all sources (red) as a function of the PL index $a$, considering an observation strategy consistent with CHIME's 1st catalog. We assume $\mathcal{R}_{\rm 0}(>10^{26}\mbox{erg Hz}^{-1})=5\times 10^{-6}\mbox{hr}^{-1},E_{\rm c}=3\times 10^{33}\mbox{erg Hz}^{-1},\gamma=1.7$ as inferred from observations. Shaded regions depict the inferred ranges of these properties from observations. Solid lines are the results of numerical integration (Eq. \ref{eq:Nkdetfull}). The model matches observations well within the shaded yellow range.}
\label{fig:Numerical}
\end{figure*}

We take a distribution of source rates $P(\geq \mathcal{R}_{\rm *})=(\mathcal{R}_{\rm *}/\mathcal{R}_{\rm 0,*})^{-a}$ with $a>0$, so that observed CHIME bursts might be dominated by a rare but prolific population of FRB repeaters (see \S \ref{sec:varyrate}).
The ratios $N_{\geq 1}/N_{\geq 0}, N_{\geq 2}/N_{\geq 1}$ depend on the value of $a$
\begin{eqnarray}
\label{eq:N2N1}
 {\small    \frac{N_{\geq 1}}{N_{\geq 0}}\!\approx\!\left\{ \begin{array}{ll}\!\lambda_{\rm lim}^{a}& 0\!<\!a\!<\!1 \\ \!\lambda_{\rm lim} & a>1\ .
\end{array}\right. ;
    \frac{N_{\geq 2}}{N_{\geq 1}}\!\approx\!\left\{ \begin{array}{ll}\!2^{-a}& 0\!<\!a\!<\!1 \\
\!\frac{\lambda_{\rm lim}^{a-1}}{2^{a}} & 1\!<\!a\!<\!2 \\ \!\frac{\lambda_{\rm lim}}{4} & a>2\ .
\end{array}\right.}
\end{eqnarray}
and $\lambda_{\rm lim}\approx \lambda_{\rm 0,lim}$ as before.
For $a<1$, the fraction of repeaters is too large $N_{\geq 2}/N_{\geq 1}\geq 2^{-a}>1/2$, while for $a>2$, the estimate in Eq. \ref{Eq:NSrc_sing} is approximately applicable, and once more the total number of sources is much too low. However, for intermediate values of $1<a<2$, and specifically, for $a-1\ll 1$, one finds that $N_{\geq 1}/N_{\geq 0}\approx \lambda_{\rm 0,lim}\ll N_{\geq 2,\rm CH}^{\rm all-sky}/N_{\geq 1,\rm CH}^{\rm all-sky}$.
Indeed taking $a\approx 1.13$ simultaneously reproduces all three observables. As mentioned above, the simple analytic estimate needs to be adjusted to account for a few effects: (i) A more realistic, not broken PL approximation of $N_{\geq k}(>\Phi)$, (ii) cosmological distance, time dilation and frequency redshift corrections, (iii) The average spectrum of FRB sources, suggesting that bursts with a given spectral flux are less common at high frequencies, (iv) the non-uniform observation time spent by a large sky survey per point in the observed sky and (v) the number density of sources which in general depends on redshift. These factors are all accounted for by direct numerical integration in \S \ref{sec:MonteCarlo}. The results are shown in Fig. \ref{fig:Numerical}. As seen in the figure, the analytic result quoted above is well reproduced by this more precise calculation. The results of the latter, indicate that $1.1\lesssim a\lesssim 1.3$  can simultaneously match the observed $N_{\geq 1}/N_{\geq 0}$, $N_{\geq 2}/N_{\geq 1}$ and $\lambda_{\rm 0,lim}$.

An alternative possible resolution is to assume a connection between source rate and maximum burst energy, the results corresponding to this situation are described in \S \ref{sec:correlationb}.

\section{Implications for magnetar sources}
\label{sec:magnetar}
Our analysis outlined in \S \ref{sec:Infer} and appendix \ref{sec:MonteCarlo}, is model independent and does not rely on FRBs being associated with a specific source type or emission mechanism. Nonetheless, various lines of evidence point towards magnetars being the underlying sources of a large fraction, and perhaps all of FRBs. In particular, Eq. \ref{FRB-rate} that has been applied here to the rate of FRBs per source, has also been shown to hold across the Galactic magnetar population, and across temporally removed burst episodes in individual magnetars (see e.g. \citealt{1996Natur.382..518C,Gogus1999,Gavriil04}). Moreover, these studies have found the PL index to be $\gamma\approx1.7$, consistent with the values inferred for FRBs\footnote{Magnetar bursts are thought to be associated with crust dislocations \citep{TD2001}.
If, with more observations, it is established that the value of $\gamma$ for FRBs is slightly different than that for magnetar bursts, this could indicate a difference in the scaling of X-ray/$\gamma$-ray bursts and that of FRBs with respect to the underlying distribution of dislocation amplitudes (see details in \citealt{Wadiasingh2020}).}. Other lines of evidence include radiation properties (coherent emission, rapid variability, narrow spectral width, bright luminosity, see \citealt{Popov_2010,Kumar+17,BK2020}) as well as FRB host galaxies, offsets relative to their hosts and burst statistics \citep{Wadiasingh2019,Cheng+20,Gordon2023,Totani2023}. The most direct association comes from the discovery of FRB 20200428, originating from a well-known Galactic magnetar \citep{chimeSGR,STARE2020}. However, this is also a unique source, in that its rate (at a given energy) is very low and the number density of similar sources in the Universe is very large (hence its location in the top left of figure \ref{fig:ratevsn}).
This then raises the question of whether magnetars can potentially accommodate the entire range of FRB source activity levels.

As described in \S \ref{sec:Infer}, when calibrating our framework by the FRB rate per source and volumetric density of Galactic magnetars, we are able to simultaneously reconcile the ratio of observed non-repeaters compared to all SGRs, the ratio of observed repeaters to non-repeaters and the typical distances of detected repeaters, being closer by a few tens of percent than those of detected non-repeaters. This is done for a population with a continuous PL function describing the probability of the source having a given activity level ($P(\geq \mathcal{R}_*)\propto \mathcal{R}_*^{-a}$, see Eq. \ref{eq:a_def}). Remarkably, once calibrating the rate and density of the most common sources as described above, the only free parameter of the model is $a$, and the range found to reproduce the observables is obtained for the same value of $a$ that is independently obtained from observations of different individual sources (see figure \ref{fig:ratevsn} and appendix \ref{sec:figure1_explained}). This argues in favor of magnetars being the source of a large fraction (or all) of the FRB population.

For $\gamma<2$, inferred for FRB sources and magnetar X-ray / $\gamma$-ray bursts, the emitted power per source (averaged over sufficiently long timescales) is dominated by bursts with energies comparable to the maximal energy, $E_{\rm c}$. The (isotropic equivalent) radio energy released per source during an activity time $t$ is given by $E_{\rm rel}\approx 6\times 10^{36}(\mathcal{R}_*/\mathcal{R}_{0,*})(t/30\mbox{ yr})(E_{\rm c}/E_{\rm lim})^{2-\gamma} \mbox{ erg}$. Since we infer $|a-1|\ll 1$, there is large probability of detecting sources with $\mathcal{R}_{*}\gg \mathcal{R}_{0,*}$ (despite their intrinsic rarity) and indeed such sources should dominate the population of CHIME repeaters (see Fig. \ref{fig:Nphivarrate}). As shown in Fig. \ref{fig:ratevsn}, for FRB 20121102A-type sources, $\mathcal{R}_*/\mathcal{R}_{0,*}\approx 3\times 10^8$. This leads to $E_{\rm rel}\approx 1.8\times 10^{45}$ for the same beaming factor, radio efficiency and activity time, as consistent with the independent estimate $E_{\rm rel}\approx 8.4\times 10^{44}\mbox{ erg}$ done in \cite{BK2025}\footnote{the small difference arises mostly due to uncertainty in the assumed spectral index and $\gamma$, which may also vary somewhat between FRB 20121102A and the average FRB population}. For an extremely active source such as FRB 20121102A, \cite{BK2025} have shown that the magnetic energy reservoir becomes constraining and special conditions have to be met to satisfy it (e.g. regarding the alignment of magnetic, rotation and viewing angle axes as well as the age of the source and its internal magnetic field strength). The apparent source density of FRB 20121102A-like sources is extremely small, $\sim 10^{-9}$ that of standard SGRs (see figure \ref{fig:ratevsn}). This can be a combination of intrinsic magnetar properties (e.g. extremely strong and rare magnetic field strength or very young age) as well as strong beaming (e.g. due to alignment of the magnetic ans spin axes, see \citealt{BK2025}). We return in more detail to the potential role of the latter in \S \ref{Sec:conclude}.

\section{Conclusions}
\label{Sec:conclude}
We have investigated the hypothesis that all FRBs originate from repeating sources. While the intrinsic repetition rates of individual sources at a fixed energy, $\mathcal{R}_*$, span a wide range, we find that the inferred number density of similarly active sources, $n_{\rm obs}$  varies correspondingly. Remarkably, these two quantities follow a simple inverse relationship, well described by a Zipf-like law: $n_{\rm obs}\approxprop \mathcal{R}_*^{-1}$, extending across approximately 8–9 orders of magnitude.
Zipf’s law—where frequency scales inversely with rank—has been observed in numerous natural systems. In astrophysics, similar patterns have been noted in the distributions of solar flares \citep{1991ApJ...380L..89L} lunar crater sizes \citep{1994hdtc.conf..359N}, and the size distributions of both cosmic voids \citep{2005EPJB...47...93G} and superclusters \citep{2021A&A...651A.114D}.

The applicability of Zipf’s law to FRBs is derived directly from observational data. It implies that the effective distribution of intrinsic repetition rates (above a fixed energy) follows a PL: $P(>\mathcal{R}_*)\propto \mathcal{R}_*^{-a}$ with $a$ close to unity. 
We explored the implications of such a distribution for the observable properties of FRBs detected in wide-field surveys such as ASKAP, DSA, and CHIME. Specifically, we computed the expected distributions of fluence, distance, and observed repetition number as functions of $a$ and $\gamma$ (where $\gamma$ is the PL index describing the burst energy distribution from individual sources).
Our analysis shows that the behavior of the FRB population can be qualitatively divided into four distinct regimes in the $a-k$ parameter space (with $k$ denoting the number of observed bursts from a source), depending on the relationship between $a,k$, and the critical value $a_{\gamma}=k_{\gamma}=3/(2(\gamma-1))$.
For typical observed values - $\gamma\approx1.7$ and $a\approx 1$ - repeaters are dominated by a small number of highly active sources, while non-repeaters include comparable contributions from both rare, active sources and more common, intrinsically inactive ones. All observed FRBs tend to originate from distant sources, although repeaters are, on average, slightly closer than non-repeaters (consistent with trends in dispersion measures; see Fig. \ref{fig:distanceandrate}).
A much sharper distinction exists between sources that are observed at least once and those that are never detected. The latter are predicted to be far more numerous and significantly less active. For example, the corrected repetition rate of SGR 1935+2154 (accounting for the observed energy of its bursts) yields an expected detection rate (above CHIME threshold) of only $\sim few\times 10^{-9}$ bursts per such source at cosmological distances - comparable to the inferred ratio of CHIME-detected sources to the estimated number of inactive, SGR-like sources in the Universe. This apparent coincidence emerges naturally if the source repetition rate distribution follows a power law with index $a\approx 1.1-1.3$. Notably, the same distribution also reproduces the observed ratio between CHIME repeaters and apparent non-repeaters. Since $a-1\ll 1$, the relative fraction of repeaters increases only very slowly with increased observation time or improved sensitivity (as $N_{\geq 2}/N_{\geq 1}\propto (\Phi_{\rm lim}^{1-\gamma}T)^{a-1}$ see \S \ref{sec:MonteCarlo} and Fig. \ref{fig:distanceandrate}). This weak dependence could be misinterpreted as evidence that repeaters and non-repeaters represent fundamentally different source populations, even if all FRBs originate from the same underlying class.

A key strength of our analysis is its independence from any specific progenitor or emission model. Nonetheless, the consistency of the least active FRB sources with known Galactic SGRs, combined with the success of a single PL of the activity-rate distribution in explaining the full observed population, naturally points to magnetars as the dominant—or possibly sole—origin of FRBs. What then accounts for the wide variation in inferred source densities and activity levels among different magnetars? 
\cite{BK2023,BK2025} suggest that the inclination between spin and magnetic axes ($\alpha$) can play a major role in understanding various apparent differences between repeaters and non-repeaters (including rate, source densities and energetics as relevant for the present analysis, but also resolving the lack of rotational periodicity in arrival time data of active repeaters, explaining the lack of polarization angle swing in active repeaters compared to polarization angle swing seen in some non-repeaters and providing a geometrical explanation for the CHIME spectro-temporal dichotomy). This raises the question - what fraction of the spread in inferred source densities can be accounted for by magnetic inclination? The fraction of the sky that is illuminated by a given misaligned NS's polar cap is $\propto \sin \alpha\approx \alpha$. The intrinsic probability of the inclination being $\leq \alpha$, $P_{\rm int}(<\alpha)$, depends on the initial distribution and on potential alignment that may take place as the magnetar evolves, both of which are uncertain. Therefore, all else being equal, the probability to observe a NS with inclination smaller than $P_{\rm ob}(<\alpha)$ is $\sim P_{\rm int}(<\alpha)\alpha$. An illustrative example is the case of a purely isotropic distribution. In this case we have $P_{\rm int}\propto \alpha^2$ and $P_{\rm ob}(<\alpha)\propto \alpha^3$. This holds for $\rho\lesssim \alpha\lesssim \pi/2$ where $\rho$ is the half opening angle of the polar cap in the observer frame. Considering likely values of the spin period, one might expect $10^{-2.5}\lesssim \rho \lesssim 10^{-1.5}$. Therefore, it is plausible that $\alpha^3$ varies by up to $\sim 7.5$ orders of magnitude and that inclination plays a significant and perhaps even the major rule in explaining the inferred source density variations. The rotation of the emission beam also affects the observed rate, which will increase with decreasing $\alpha$ according to $\mathcal{R}_{\rm ob}\propto \alpha^{-1} \mathcal{R}_{\rm int}(\alpha)$, where $\mathcal{R}_{\rm int}(\alpha)$
is the intrinsic dependence of the rate on $\alpha$. Agreement with the inferred $a\approx 1.1-1.3$ requires (for an isotropic inclination distribution) that $\mathcal{R}_{\rm int}(\alpha)\approxprop \alpha^{-1.8}-\alpha^{-1.3}$. Whether or not this can be reconciled depends on uncertain bursting properties and is an interesting topic for future investigations. Variation in other properties such as magnetic field strength or age-as well as temporal variations in burst rate, energy distribution slope $\gamma$, or maximum burst energy-can also partially contribute to the intrinsic rate - source density Zipf law. Finally, an alternative possibility relates to the presence of a rare sub-population of ultra-long-period magnetars that might be especially prolific FRB emitters (see e.g. \citealt{Beniamini+20, Beniamini2023}). Note that in this case, the polar cap angle can be much smaller than mentioned above, making it easier to explain a large range in rates and source densities, even without a strong dependence of $P_{\rm int}(<\alpha),\mathcal{R}_{\rm int}(\alpha)$ on $\alpha$. 

\bigskip\bigskip
\noindent {\bf ACKNOWLEDGEMENTS}

\medskip
We thank Zorawar Wadiasingh, Om Gupta, Rob Fender and the anonymous referee for helpful comments on the manuscript.
The work was funded in part by an NSF grant AST-2009619 (PK), a NASA grant 80NSSC24K0770 (PK and PB), a grant (no. 2020747) from the United States-Israel Binational Science Foundation (BSF), Jerusalem, Israel (PB) and by a grant (no. 1649/23) from the Israel Science Foundation (PB).

\appendix
\restartappendixnumbering

\section{Numerical verification}
\label{sec:MonteCarlo}
To verify the validity of our simplifying assumptions presented in the main text, we have carried also a numerical calculation in which we account for appropriate redshift corrections, as well for a realistic distribution of observation time per location on the sky.

In particular, we assume the distribution of FRB sources, $n_{\rm src}(z)$ is proportional to the cosmic SFR \citep{Gupta2025},
\begin{equation}
    n_{\rm src}(z)=A_{\rm src}^{\rm sfr}\dot{m}_{\ast}(z) = A_{\rm src}^{\rm sfr}
    \begin{cases}
        0.015 \dfrac{(1+z)^{2.73}}{1 + [(1+z)/3]^{6.24}} & \text{if } 0 < z < 4,\\
        10^{-0.257z - 0.275} & \text{if } 4 \leq z \leq 14.
    \end{cases} M_{\odot}\mbox{ yr}^{-1}\mbox{ Mpc}^{-3}.
    \label{eq:sfrd}
\end{equation}
where $A_{\rm src}^{\rm sfr}$ is the conversion fraction from star formation rate to population density of FRB sources. If the latter are magnetars, then approximately, $A_{\rm src}^{\rm sfr}\approx 20 \mbox{yr} M_{\odot}^{-1}$ (corresponding to roughly 30 magnetars in the Milky Way and a number density of Milky-Way like galaxies of $\sim 0.01\mbox{ Mpc}^{-3}$).
Some recent studies \citep{Qiang2022,ZhangZhang2022,Lin2024,Gupta2025,HorowiczMargalit2025,Acharya2025} have suggested that FRBs arising from a mix of SFR and stellar mass following populations, fit better with the observed redshift and host galaxy distributions. Such a possibility is straightforward to include in this analysis. Since its effect on the results of this study are minor and in order to avoid introducing additional weakly constrained parameters, we ignore this complication in the estimates below. 
We assume also that the average spectrum of FRBs can be characterized using $E_{\nu}\propto \nu^{\alpha}$. 
The measured fluence of an FRB is then given by $\Phi=E_{\nu_0}(1+z)^{\alpha}/(4\pi r^2)$ where $\nu_0$ is the frequency at which the fluence is measured, $r$ is the proper distance and the factor $(1+z)^{\alpha}$ is a $k$-correction, i.e. converting from the measured $\nu_0$ to the corresponding frequency in the comoving frame $\nu=\nu_0(1+z)$.

We consider a non-uniform distribution of observation time per location on the sky viewed by the survey, $d\mbox{P}/dT$. For our numerical calculations we use the distribution reported in the first CHIME catalog \citep{CHIME_1st_cat}. This corresponds to a median observation time of $\sim 20$\,hrs, but with some locations observed for much longer times.

Taking all these ingredients together, we may calculate the number of bursts detected $k$ times during a survey
\begin{eqnarray}
\label{eq:Nkdetfull}
 &   N_{\geq k}=\frac{\Omega_{\rm srv}}{4\pi}\int_0^{\infty}dT \int_{\mathcal{R}_{\rm 0,*}}^{\infty} d\mathcal{R}_{\rm *}\int_0^{\infty} dV(z)  \frac{d\mbox{P}}{dT} \frac{d\mbox{P}}{d \mathcal{R}_{\rm *}} n_{\rm src}(z) \mbox{Pois}\left(k\bigg\vert \frac{T}{1+z}\mathcal{R}_{\rm *} \left[\frac{E_{\rm lim}(z)}{E_{\rm *}}\right]^{1-\gamma}\right)\Theta(E_{\rm c}-E_{\rm lim}(z))
\end{eqnarray}
where $\Omega_{\rm srv}/4\pi$ is the fraction of the sky viewed by the survey (approximately 0.5 for CHIME), $\Phi_{\rm lim}$ is the limiting (spectral) fluence of the survey and $E_{\rm lim}(z)=4\pi r^2\Phi_{\rm lim} (1+z)^{-\alpha}$. $d\mbox{P}/d\mathcal{R}_{\rm *}\propto \mathcal{R}_{\rm *}^{-a-1}$ is the probability of a source having a given intrinsic rate (at $E_{\rm *}$). Pois$(k\vert\lambda)$ is the Poisson probability distribution of observing $k$ events considering an expected value of $\lambda$. The value $\lambda=\frac{T}{1+z}\mathcal{R}_{\rm *} \left[\frac{E_{\rm lim}(z)}{E_{\rm *}}\right]^{1-\gamma}$ accounts for the rate of the source above the minimum energy detectable by the survey and considering the (comoving) time that it has been observed by the survey.

\begin{figure}
\centering
\vspace{-5mm}
\includegraphics[width=0.34\textwidth]{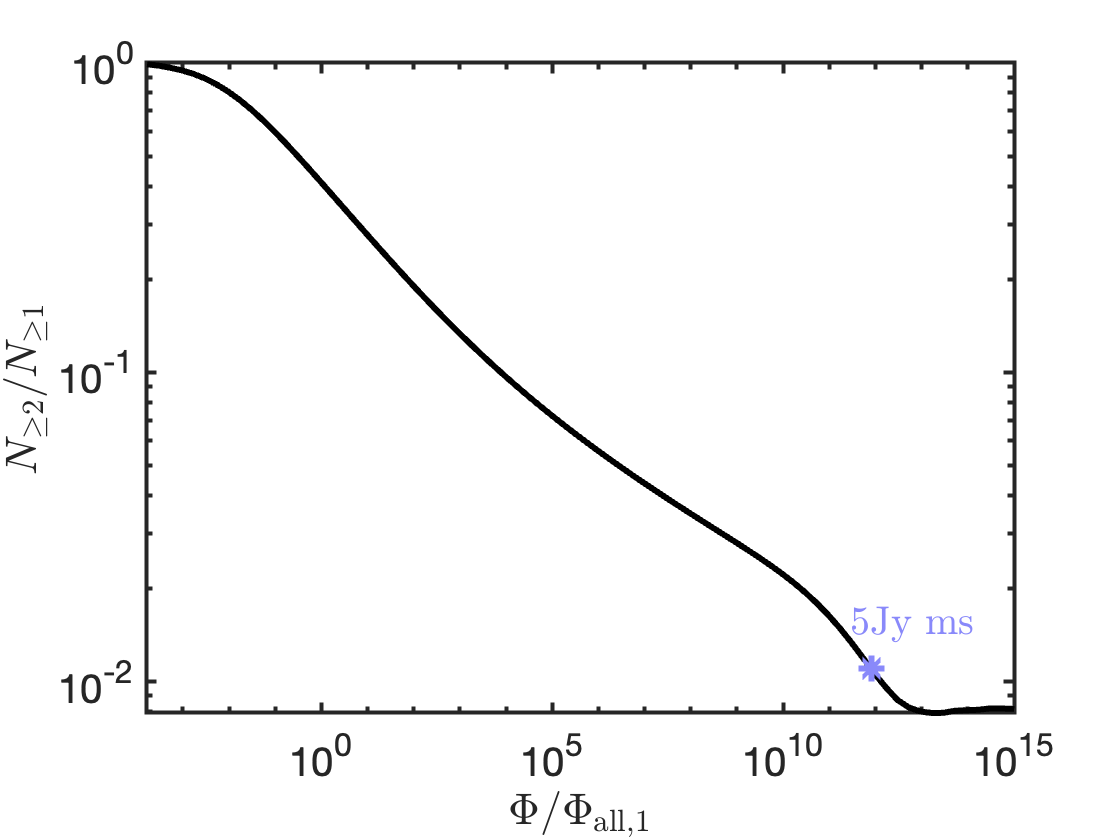}
\vspace{-1mm}
\hskip -2ex
\includegraphics[width=0.34\textwidth]{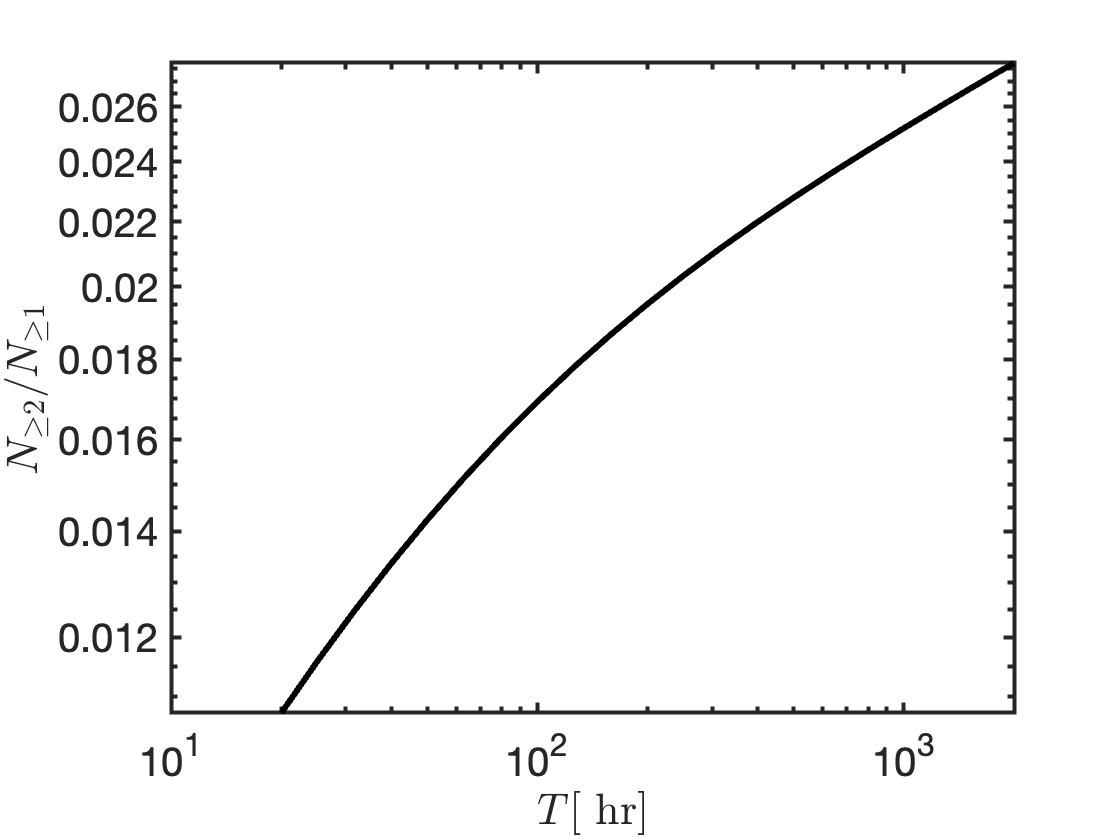}
\vspace{-1mm}
\hskip -2ex
\includegraphics[width=0.34\textwidth]{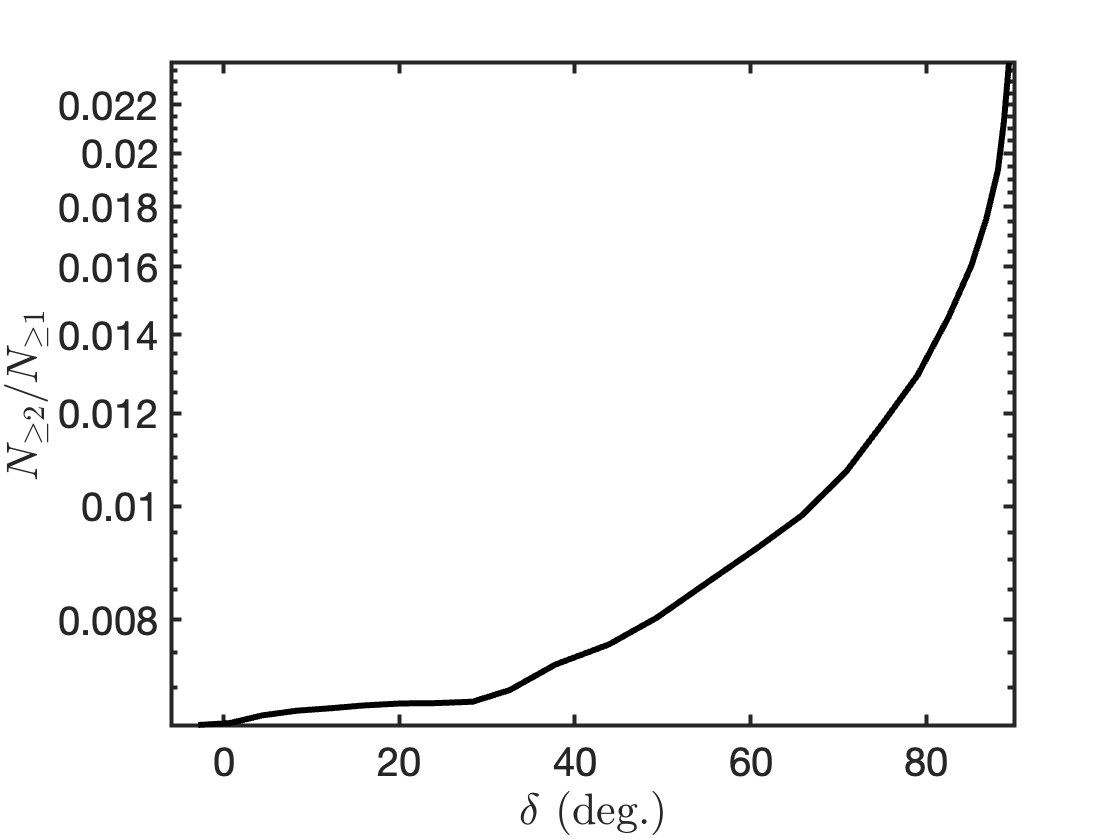}
\caption{Fraction of observable repeaters as a function of the limiting fluence (left) and as a function of mean observation time per sky location (center) using a fluence threshold of 5Jy ms (marked by an asterisk on the left panel). In both cases we adopt $a=1.1$, consistent with observational constraints (see Fig. \ref{fig:Numerical}). The repeater fraction is very weakly dependent on either $T$ or $\Phi$. The right panel shows how the repeater fraction changes with declination assuming the distribution of exposure vs. declination from the first CHIME catalog.}
\label{fig:N2N1vsPhiandT}
\end{figure}

\begin{figure}
\centering
\vspace{-5mm}
\includegraphics[width=0.34\textwidth]{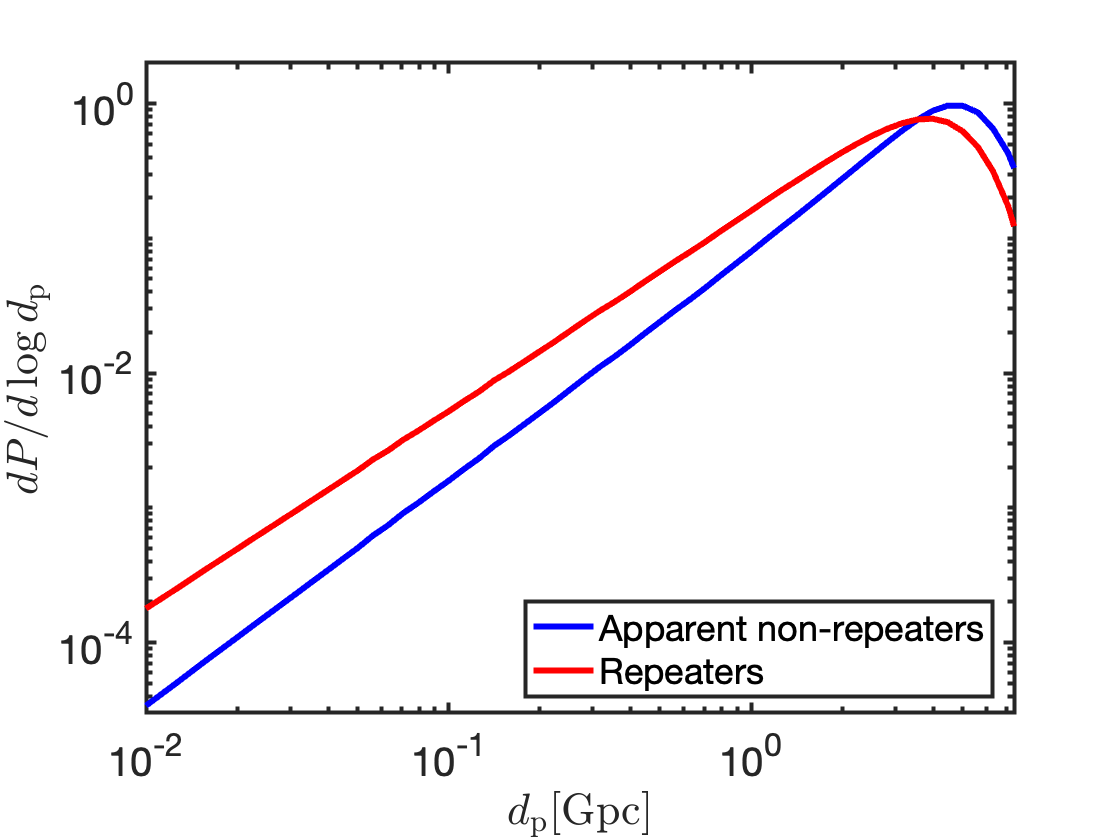}
\vspace{-1mm}
 \hskip -2ex
\includegraphics[width=0.34\textwidth]{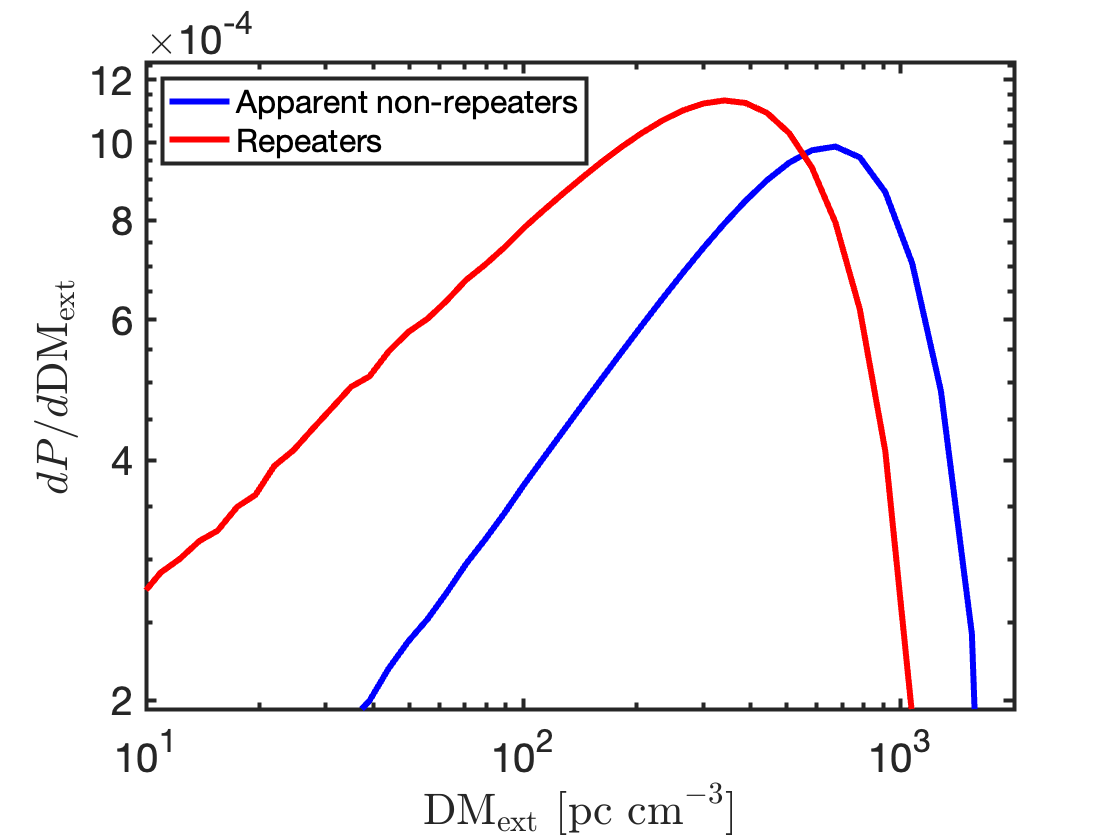}
\vspace{-1mm}
 \hskip -2ex
\includegraphics[width=0.34\textwidth]{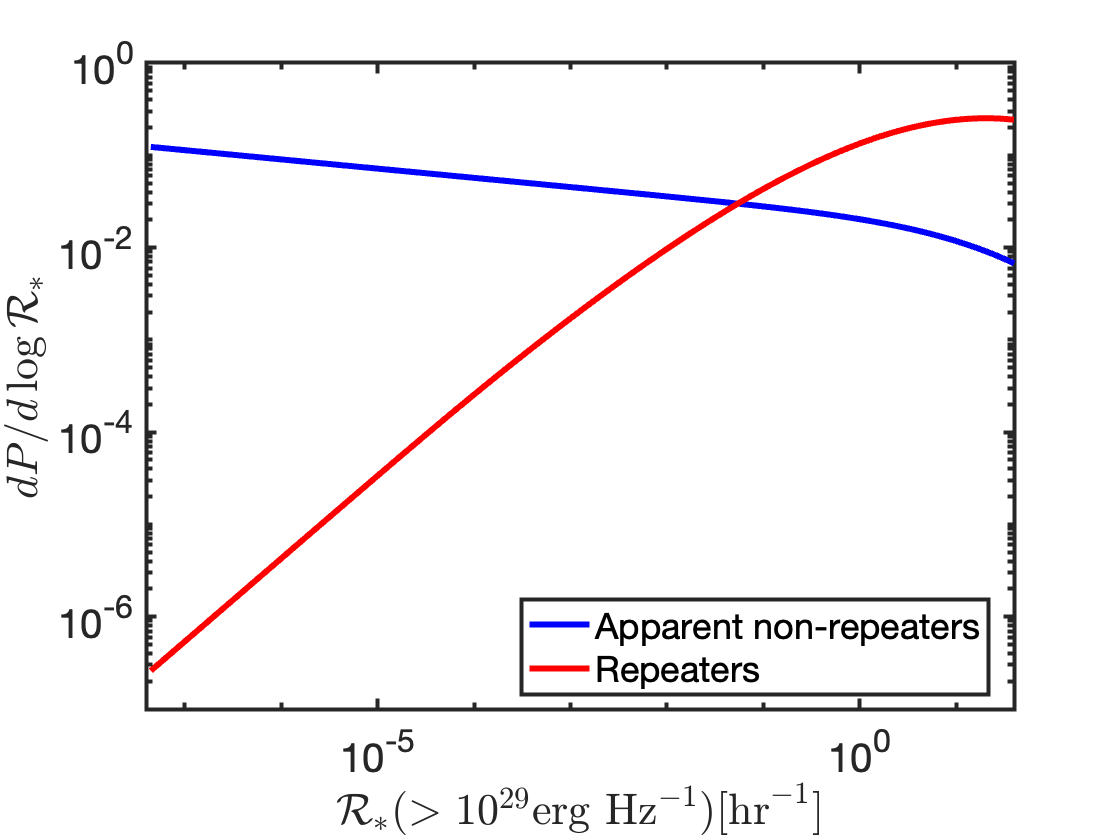}
\caption{Distance (left), external DM (due to propagation in the intergalactic medium) and activity rate (right) distributions of repeaters and non-repeaters. Results are shown for $a=1.1$ and other parameters as in described in appendix \ref{sec:MonteCarlo} and Fig. \ref{fig:N2N1vsPhiandT}. For DM$_{\rm ext}$ we use DM$_{\rm ext}=855z\mbox{pc cm}^{-3}$, which is applicable for the $z\lesssim 3$ population that dominates the CHIME detected population.}
\label{fig:distanceandrate}
\end{figure}

The results of numerical integration for $N_{\geq 2}/N_{\geq 1}, N_{\geq 1}/N_{\geq 0}$ as a function of $a$ were shown in Fig. \ref{fig:Numerical} where it was shown that $1.1\lesssim a \lesssim 1.3$ can simultaneously match the observed $N_{\geq 1}/N_{\geq 0}$, $N_{\geq 2}/N_{\geq 1}$ and $\lambda_{\rm 0,lim}$. Fig. \ref{fig:N2N1vsPhiandT} shows how the ratio of repeaters to total detected sources, $N_{\geq 2}/N_{\geq 1}$, evolves for a fixed value of $a=1.1$ , as a function of decreasing the fluence threshold or increasing the average observation time per sky location. 
For a small value of $a-1$, the analytic expression $N_{\geq 2}/N_{\geq 1}\propto (\Phi_{\rm lim}^{1-\gamma}T)^{a-1}$ (Eq. \ref{eq:N2N1}) predicts that the repeater fraction depends only weakly on both limiting fluence and observing time. This weak dependence is confirmed by our numerical results: increasing the observation time by a factor of 10 leads to only a modest increase - by a factor of $\sim 1.8$ - in the fraction of repeaters. Reducing the fluence threshold by a similar factor yields an even smaller effect.
The figure also illustrates how the repeater fraction varies across the sky, based on the relation between exposure time and declination in the first CHIME catalog. Since CHIME observes higher declinations for longer, the fraction of repeaters is slightly elevated in those regions. However, this variation is relatively small, again reflecting the weak dependence of the repeater fraction on exposure (as a rough approximation, CHIME's exposure distribution roughly satisfies $T\propto (\pi/2-\delta)^{-1}$,  \citealt{CHIME_1st_cat}, leading to $N_{\geq 2}/N_{\geq 1}\propto (\pi/2-\delta)^{1-a}$). 
We also show, in Fig. \ref{fig:distanceandrate}. the distribution of distances, external dispersion measures and intrinsic rates for both observed repeaters and non-repeaters. As shown in Fig. \ref{fig:Nphivarrate}, for $a=1.1$, both repeaters and non-repeaters are predominantly detected at cosmological distances. Interestingly, repeaters are, on average, a few tens of percent closer than non-repeaters. This leads to typical intergalactic dispersion measures of $\mbox{DM}_{\rm ext,rep}\sim 400\mbox{pc cm}^{-3}$ for repeaters, compared to $\mbox{DM}_{\rm ext,NR}\sim 670\mbox{pc cm}^{-3}$ for non-repeaters—both consistent with observed values \citep{CHIME2023repeaters}. This agreement is particularly striking given that our modeling was developed independently of dispersion measure data, using a simple framework and approximate detectability criteria. While the two populations show some overlap in distance, they differ much more clearly in their intrinsic repetition rates: repeaters are dominated by the rarest and most active sources, whereas the non-repeater population includes a roughly equal mix of common, inactive sources and rare, highly active ones.

\section{Inferred densities and source activity rates}
\label{sec:figure1_explained}
The nearest source that is a member of a given class of objects and is within the field observable by the telescope, would be the brightest and easiest to detect\footnote{Beaming corrections can be important in order to translate the inferred source densities derived in this way to true space densities, but since it is apriori unknown, it is more robust to report the density of sources that are beamed towards the observer}.
Therefore, if the closest known member of a class of sources is at a distance $d_{\rm p}$ (for cosmological sources, $d_{\rm p}$ here is the proper distance), then the volume within which there is one source is $V_{\rm obs}=\Omega_{\rm srv} d_{\rm p}^3/3$ (where $\Omega_{\rm srv}$ is the solid angle covered by telescopes that could have detected such sources, and $\Omega_{\rm srv}\approx 2\pi$ for CHIME). The mean number of sources in the volume is given by $\lambda_{\rm obs}=V_{\rm obs}n_{\rm obs}$ where $n_{\rm obs}$ is the volumetric density of sources.
From Poisson statistics, the probability of observing one source in this volume, given $\lambda_{\rm obs}$, is $P(1|\lambda_{\rm obs})=\lambda_{\rm obs}\exp{[-\lambda_{\rm obs}]}$. Using Bayes' theorem, this can be converted to the probability that the mean number of sources in the volume is $\lambda_{\rm obs}$ given that one source was detected within the volume,
\begin{equation}
\label{eq:Bayes}
    \frac{dP(\lambda_{\rm obs}|1)}{d\lambda_{\rm obs}}\propto P(1|\lambda_{\rm obs})\frac{dP}{d\lambda_{\rm obs}}.
\end{equation}
To progress from here, one needs to make a choice regarding the prior of $dP/d\lambda_{\rm obs}$. We take $dP/d\lambda_{\rm obs}\propto \lambda_{\rm obs}^{-1}$ such that $P(<\lambda_{\rm obs})\propto \log(\lambda_{\rm obs})$ and no apriori bias is introduced in favor of large or small values of $\lambda_{\rm obs}$ (see e.g. \citealt{LBK2022} and references therein). Plugging this in Eq. \ref{eq:Bayes}, switching variables from $\lambda_{\rm obs}$ to $n_{\rm obs}$ and integrating we get
\begin{equation}
\label{eq:Bayes2}
    P(<n_{\rm obs}|1)= 1-\exp[-V_{\rm obs}n_{\rm obs}].
\end{equation}
Eq. \ref{eq:Bayes2} shows that the median value of $n_{\rm obs}$ is $0.7V_{\rm obs}^{-1}$ and that the 90\% confidence intervals are $n_{\rm obs}\in [0.051,3]V_{\rm obs}^{-1}$. We use these values to estimate the volumetric densities of different FRB source types in figure \ref{fig:ratevsn}.

To calculate the rates of different sources, we have collected data reported in the literature. Since the activity rates of FRB sources may vary significantly over time, we have chosen, wherever possible, the observations corresponding to the longest monitoring of any given source. Errors due to time variation are calculated assuming Poisson statistics. Furthermore, since rates of sources depend on both burst energy and observed frequency, we present all rates in figure \ref{fig:ratevsn} at a common energy ($10^{28}\mbox{erg Hz}^{-1}$) and frequency (600MHz - the midpoint of the CHIME band). When observations at those energy and frequency are not available, we take the closest observed energy/frequency and extrapolate using Eq. \ref{FRB-rate} for the energy and  the `average' spectrum $E_{\nu}\propto \nu^{\alpha}$. This `averaged spectrum' takes into account that there is large variation in individual FRB spectra, some being narrow band, some broad band and some spectrally complex, but as shown by e.g. \cite{Macquart2019}, when averaged over a larger sample, the fluence is reasonably represented by a power-law behavior. Using the average spectrum and energy dependent rate, if we know the rate of bursts with energy $>E_1$ at some $\nu_1$, $\mathcal{R}(>E_1|\nu_1)$, then the source would on average have the same rate of bursts at $\nu_0$ and above $E_0=E_1(\nu_0/\nu_1)^{\alpha}$ (this is straightforward for wide band bursts - each burst with energy $E_1$ at $\nu_1$ has an energy $E_0$ at $\nu_0$. It is therefore the most self-consistent definition of $\alpha$ that guarantees the results do not depend on whether individual bursts are wide or narrow band). With this assumption, we can get the rate of bursts above any other value of $E_{\nu}$ via 
\begin{equation}
\label{eq:ratewithfreq}
  \mathcal{R}(>\!E_{\nu}|\nu_0)=\mathcal{R}(>\!E_0|\nu_0)\bigg(\frac{E_{\nu}}{E_0}\bigg)^{1-\gamma}=\mathcal{R}(>\!E_1|\nu_1)\bigg(\frac{E_{\nu}}{E_0}\bigg)^{1-\gamma}=\mathcal{R}(>\!E_1|\nu_1)\bigg(\frac{E_{\nu}}{E_1}\bigg)^{1-\gamma}\bigg(\frac{\nu_1}{\nu_0}\bigg)^{\alpha(1-\gamma)}
\end{equation}
When observations of a given source are not available at 600 MHz and $10^{28}\mbox{erg Hz}^{-1}$, Eq. \ref{eq:ratewithfreq} is used to extrapolate the rates. The errors are extrapolated in the same way, and in addition we allow for uncertainties in the average values of $\gamma$ and $\alpha$. Unless, reported otherwise for a specific source, we take $\gamma=1.7\pm 0.2$ and $\alpha=-1.5\pm 0.3$. Uncertainties in figure \ref{fig:ratevsn} mark 90\% confidence intervals. We generally choose observations at energies / frequencies as close as possible to our reference values, in order to minimize errors due to the extrapolation.

We include in Fig.\ref{fig:ratevsn} a calculation of the equivalent volumetric source density and rate that would be associated with CHIME detected sources under the assumption (disfavored in this work) that there is a single activity rate per source underlying all sources (as per \S \ref{sec:Singlesource}).
These are calculated in the following way. 
From the number of sources detected twice and those detected once, we can calculate the number of all sources (i.e. including those which have not been detected at all). This can be done in an approximate way using Eq. \ref{eq:repratio} and correcting for the fraction of sky visible by CHIME (see, Eq. \ref{Eq:NSrc_sing}), or in a slightly more accurate way using Eq. \ref{eq:Nkdetfull} with $dP/d\mathcal{R}_*=\delta(\mathcal{R}_*-\mathcal{R}_{0,*})$. Once the total number of sources, $N_{\geq 0}^{\rm sgl}$, has been calculated, we use Eq. \ref{eq:sfrd} to find the corresponding source density at $z=0$. Dividing the all-sky equivalent rate of bursts discovered by CHIME (as reported in the 1st catalog, \citealt{CHIME_1st_cat}) by $N_{\rm \geq 0,CH}^{\rm all-sky}$ we obtain the average rate per source under these assumptions. The latter is then extrapolated from $E_{\rm lim}$ to $10^{28}\mbox{erg Hz}^{-1}$ using Eq. \ref{FRB-rate} and $\gamma=1.7\pm0.2$ as described in the previous paragraph. 
We emphasize that these ``effective" densities and rates are derived using a proof by contradiction approach: they assume that all the different sources observed by CHIME can be described by a single characteristic activity rate and typical source density. However, as illustrated by the wide spread of black points in the parameter space in Fig. \ref{fig:ratevsn}, this assumption is clearly inconsistent with the data based on the calculation of the rates and source densities of individually well monitored sources.

While using the technique outlined at the top of this appendix, $n_{\rm obs}$ is only well estimated by the nearest member of a given observationally defined sub-class of sources (e.g. one with a given repetition rate or other clearly identifiable observable distinctions), the individual source distances ($d_{\rm p}$) maintain their physical meaning also for non-nearest members. An additional advantage of considering the source distances rather than source densities is that, as explained above, in order to estimate $n_{\rm obs}$, it is crucial to have an estimate of the sky coverage of the detecting telescope in the observing mode that detected the burst. This issue is particularly important for non-repeaters detected by instruments other than CHIME, which typically probe only a small fraction of the sky (leading to a large and uncertain correction factor in any attempt to estimate source densities). Relaxing the requirement to have reliable and physically meaningful estimates of $n_{\rm obs}$, allows us to expand the sample to include additional bursts, FRBs: 20180924, 20190711, 20190110C, 20190425A, 20190520B, 20190303A, 20191106A, 20200223B with well constrained rates and distances \citep{James2020,Kumar2021,Niu2022,Uno2025,Chawla2025,Shannon2025}.
In the bottom panel of Fig.\ref{fig:ratevsn} we plot the rate per source ($\mathcal{R}$) directly as a function of their distance ($d_{\rm p}$) without attempting to infer densities from non-nearest members or from surveys with unreported sky coverage. Consistency with the PL distribution of activity rates, $n_{\rm obs}\propto \mathcal{R}^{-a}$ is equivalent to an upper limit in the $d_{\rm p}-\mathcal{R}$ plane in which sources are found, which scales as $\mathcal{R}\propto d_{\rm p}^{3/a}$ (by construction, individual sources can be further away from the nearest member of their source sub-class, but not vice versa).

\section{Correlation between rate and maximal energy}
\label{sec:correlationb}
The activity rate of sources (at some fixed energy) may be correlated with the maximum burst energy. This is a natural expectation of at least some relevant physical scenarios. For example, for magnetar sources, both quantities are likely positively correlated with the magnetar's internal field strength \citep{Beniamini2025}.
We assume a PL relation, $E_{\rm c}=E_* (\mathcal{R}_{*}/\mathcal{R}_{0,*})^b$, where $\mathcal{R}_{0,*}$ is the rate of bursts (per source) at $E_*$ for sources for which $E_*$ is the maximum burst energy and $\mathcal{R}_*$ is the rate of bursts per source at the same energy for sources for which $E_{\rm c}$ is the maximum burst energy. 
It is convenient to choose $E_*$ as the maximum energy of the most common (and least active sources). $b=0$ corresponds to the case discussed in \S \ref{sec:varyrate}. Instead, for $b>0$, if $E_{\rm lim}>E_*$, then detection of bursts from sources for which $E_{\rm lim}>E_{\rm c}$ is suppressed. This leads to a minimum rate at $E_*$, $\mathcal{R}_{*}=\mathcal{R}_{0,*}(E_{\rm lim}/E_*)^{1/b}$. Using $P(>\mathcal{R}_*)\propto \mathcal{R}_*^{-a}$ we see that these sources, with $E_{\rm c}\approx E_{\rm lim}$ represent a small fraction, of order $(E_{\rm lim}/E_*)^{-a/b}$ relative to the most common, and least active sources. The mean number of detectable bursts produced by such sources is $\lambda_{\rm lim}=\lambda_{0,*}(E_{\rm lim}/E_*)^{1-\gamma+1/b}$.
There is therefore a critical value of $b=b_c\equiv 1/(\gamma-1)$, such that for $b\ll b_c$, $\lambda_{\rm lim} \gg\lambda_{0,*}$. In particular, if $\lambda_{\rm lim}>1$, this implies that most observed sources will be observed multiple times. As $b$ grows, $\lambda_{\rm lim}$ drops and so does the relative frequency of sources detected multiple times.
Moreover, since $\Phi_{\rm all,k}\propto \mathcal{R}_*^{1/(\gamma-1)}$ while $\Phi_{\rm c}\propto \mathcal{R}_*^b$ we see that for $b<b_c$ $\Phi_{\rm c}$ grows more slowly with an increasing rate than $\Phi_{\rm all,k}$. This means that there is a critical rate (or $\lambda$) for which the two become equal $\bar{\lambda}=\lambda_{0,*}^{-b(1-\gamma) \over 1-b\gamma+b}k^{1\over 1-b\gamma+b}$ and a corresponding fluence $\bar{\Phi}=\Phi_* (k/\lambda_{0,*})^{b \over 1-b\gamma+b}\gg\Phi_*$ (note that $\bar{\Phi}$ increases with $k$). Analogously to the discussion in \S \ref{sec:varyrate}, for $\Phi<\bar{\Phi}$ $N_{\geq k}(>\Phi)\propto \Phi^{a(1-\gamma)}k^{-a}$ and for $\Phi>\bar{\Phi}$, $N{\geq k}(>\Phi)\propto (\bar{\Phi}/\Phi_{\rm all,k})^{-a(\gamma-1)}(\Phi/\bar{\Phi})^{-3/2}\propto \Phi^{-3/2}k^{-a+\frac{b[3/2-a(\gamma-1)]}{1-b(\gamma-1)}}$.  Instead, for $b>b_c$, $\Phi_{\rm c}$ grows faster than $\Phi_{\rm all,k}$ and as a result $N(>\Phi)\propto \Phi^{a(1-\gamma)}k^{-a}$ holds up to arbitrarily large $\Phi$.

As explained above only a fraction $(E_{\rm lim}/E_{*})^{-a/b}$ of all sources are detectable above $\Phi_{\rm lim}$ and they are characterized by $\lambda_{\rm lim}=\lambda_{0,*}(E_{\rm lim}/E_*)^{1-\gamma+1/b}$. Requiring $N_{\geq 2}/N_{\geq 1}\ll 1$, leads to $\lambda_{\rm lim}\ll 1$ and in turn to
\begin{eqnarray}
\label{eq:N1N0}
  & \frac{N_{\geq 1}}{N_{\geq 0}}\!\approx\!\left(\frac{E_{\rm lim}}{E_*}\right)^{-\frac{a}{b}}\left\{ \begin{array}{ll}\!\lambda_{\rm lim}^{a}& 0\!<\!a\!<\!1 \\ \!\lambda_{\rm lim} & a>1\ .
\end{array}\right.\approx \lambda_{\rm 0,*}\left\{ \begin{array}{ll}\!\left(\frac{E_{\rm lim}}{E_*}\right)^{1-\gamma}& 0\!<\!a\!<\!1 \\ \!\left(\frac{E_{\rm lim}}{E_*}\right)^{\frac{1-a}{b}+1-\gamma} & a>1\ .
\end{array}\right. 
\end{eqnarray}
where $E_*$ is the maximum energy of bursts from SGR-like sources and $\lambda_{0,*}$ is the mean number of bursts produced by such a source at this typical energy and where the relation between $N_{\geq 2}/N_{\geq 1}$ and $\lambda_{\rm lim}$ is still described by Eq. \ref{eq:N2N1}.

While fixing the rate of SGRs at $10^{26}\mbox{erg Hz}^{-1}$ as above (and taking $a>1$), there is an approximate degeneracy in $E_*,b$, for a given value of $a$ (which leaves $N_{\geq 2}/N_{\geq 1}$ and $N_{\geq 1}/N_{\geq 0}$ unchanged).
This means that there are effectively only two parameters (which can be chosen as $a$ and $E_*(b)$ that are needed to uniquely determine $N_{\geq 2}/N_{\geq 1}$ and $N_{\geq 1}/N_{\geq 0}$. This degeneracy holds as long as $E_*<E_{\rm lim}\approx 1.6\times 10^{32}\mbox{erg Hz}^{-1}$.
The numerical evaluation of Eq. \ref{eq:Nkdetfull} accounting for the rate -  maximal energy correlation, give $a\approx 1.1$ and $E_*\approx 10^{32.2-3.47b}\mbox{erg Hz}^{-1}$. 

For $b>0$, sources dominating the CHIME sample, satisfy $\lambda_{\rm CH}(E_{\rm c})=\lambda_{0,*} (E_*/10^{26}\mbox{erg Hz}^{-1})^{1-\gamma}(E_{\rm c}/E_{\rm *})^{1/b+1-\gamma}\approx 1.3\times 10^{-5} (E_{\rm c}/E_{\rm lim})^{1/b+1-\gamma}$ with $E_{\rm c}\geq E_{\rm lim}$, such that the results depend on $b$ for $E_{\rm c}>E_{\rm lim}$. This is more than three orders of magnitude greater than the rate of bursts per SGR as extrapolated to $E_{\rm lim}$. We see also that the (isotropic equivalent) energy needed to power this activity is $E_{\rm rel}\approx t\mathcal{R}_{\rm CH}(E_{\rm c})E_{\rm c}\approx 1.6\times 10^{40} (t/30\mbox{ yr})(E_{\rm c}/E_{\rm lim})^{1/b+2-\gamma}\mbox{ erg}$, where $t$ is the time over which the source has been active at this level. This energy is many orders of magnitude lower than the magnetic energy reservoir of a magnetar ($E_{\rm mag}=1.6\times 10^{49}(B/10^{16})^2\mbox{ erg}$), even after accounting for the fact that typically only a small fraction of the dissipated energy may be channeled towards the radio band. Since we infer that $a-1\ll 1$, a wide range of intrinsic source repetition rates should be probed by CHIME sources. Indeed \cite{BK2025} found that for FRB 20121102A, the energy requirements result in $E_{\rm rel}\approx 8.4\times 10^{44}\mbox{ erg}$ for the same beaming factor, radio efficiency and activity time, about $3\times 10^5$ times larger than for the typical CHIME source, as calculated above.

\end{document}